\documentclass[journal=jpcc,manuscript=article]{achemso}
\usepackage{amsmath,amssymb}
\usepackage{graphicx}
\usepackage{xcolor}
\usepackage{bm}
\usepackage{nicefrac}
\usepackage{hyperref}
\usepackage{gensymb}
\usepackage{comment}
\usepackage{pdfpages}

\renewcommand{\rm}[1]{\textrm{#1}}

\usepackage{subcaption} 

\author{Shohely Tasnim Anindo}
\affiliation{Institute of Condensed Matter Theory and Optics, Friedrich-Schiller-Universit\"at Jena, Max-Wien-Platz 1, 07743 Jena, Germany}
\alsoaffiliation{Abbe Center of Photonics, Albert-Einstein-Straße 6, 07745 Jena, Germany}
\author{Daniela T\"auber}
\affiliation{Institute of Physical Chemistry, Friedrich-Schiller-Universit\"at Jena, Helmholtzweg 4, 07743 Jena, Germany }
\alsoaffiliation{Leibniz Institute of Photonic Technology, Albert-Einstein-Straße 9, 07745 Jena, Germany}
\author{Christin David}
\affiliation{Institute of Condensed Matter Theory and Optics, Friedrich-Schiller-Universit\"at Jena, Max-Wien-Platz 1, 07743 Jena, Germany}
\alsoaffiliation{University of Applied Sciences Landshut, Am Lurzenhof 1, 84036 Landshut, Germany}
\email{christin.david@haw-landshut.de}

\title{Photothermal Expansion of Nanostructures in Photo-induced Force Microscopy}

\begin{document}
\begin{abstract}
Powerful mid-infrared illumination combined with mechanical detection via force microscopy provides access to nanoscale spectroscopic imaging in Materials and Life Sciences. Photo-induced force microscopy (PiFM) employs pulsed illumination and noncontact force microscopy resulting in unprecedented spatial and high spectral resolution. The near-field-enhanced light absorption in the materials leads to thermal expansion affecting the distance-dependent weak van der Waals (VdW) force acting between the tip and the sample. We model the non-linear impact of material characteristics and surface shape on the tip-sample interaction, the heat generation from the presence of a photo-induced electric field, the associated thermal expansion under different illumination conditions including light polarization and the feedback to the dynamic tip motion due to the expansion. Comparison of the results with our experimental investigation of a polymer nanosphere shows good agreement, contributing new insights into the understanding required for a quantitative analysis of nanostructured materials imaged using PiFM.
\end{abstract}

\maketitle

\section{Introduction}
Due to their unique physical, optical, and electrical properties, nanomaterials are well suited for a wide range of applications including optical and chemical sensing, e.g., for  environmental remediation,~\cite{oaraNeumannApplication} energy conversion for next generation photovoltaics or (photo-)catalysis,~\cite{MascarettiSchirato, fusco_chapter_2024,Atwater2010} bioscience~\cite{ApplicationNanomembrane} and medical uses,~\cite{nilaPoster, joseph_nanoscale_2024} such as the treatment of cancer,~\cite{KennedyLauraApplication, Kuncic_2018} and industrial uses.~\cite{luo_intrinsic_2022} To control the scope of the functionality of the employed nanomaterials, effective characterization is important to identify mechanisms and optimize structures. Numerous studies have been conducted on the characterization of nanoparticle (NP) distributions, such as regular nanostructures using Raman spectroscopy, for example, on nanotubes,~\cite{Oener2018, David2018a} scanning electron microscopy (SEM) for optical data storage materials,~\cite{David2014b} random metal NP distributions,~\cite{David2013b} rough morphologies.~\cite{Feng2009, David2010a} Conventional imaging methods, such as transmission electron microscopy (TEM) and SEM, are frequently employed to characterize nanostructures. Using atomic force microscopy (AFM) in addition, enables the detection of the sample's surface morphology with nanoscale resolution~\cite{Marrese,Schwartz2022} evaluating its surface roughness, surface potential, and surface stickiness.~\cite{Jger2012,Shluger1994} However, conventional AFM cannot provide information on the chemical composition and molecular vibration of the samples. This limitation is resolved by additionally probing the tip-enhanced electromagnetic near-field upon visible or infrared illumination of the tip-sample region. One such approach, tip-enhanced Raman spectroscopy (TERS) has been employed for more than a decade.~\cite{hoppener_tip-enhanced_2024} The recent invention of powerful mid-infrared (mid-IR) laser sources has triggered a fast-paced development of nanoscale infrared-spectroscopic imaging methods. Mid-IR illumination has been incorporated into scattering scanning optical near-field microscopy (sSNOM)~\cite{kanevche_infrared_2021, wang_toward_2023} and several methods combining mid-IR illumination with mechanical probing of the near-field enhanced absorption have been developed, such as photo-thermal induced resonance (PTIR, also named AFM-IR),\cite{mathurin_photothermal_2022, Schwartz2022}, mid-IR photo-induced force microscopy (mid-IR PiFM, also named PiF-IR),\cite{sifat_photo-induced_2022, joseph_nanoscale_2024, davies-jones_photoinduced_2023} and mid IR peak force microscopy (PFIR).~\cite{Wang_2019_generalizedHeterodyne}
In PiFM (including PiF-IR) the tip-enhanced electromagnetic near-field is induced using a monochromatic illumination within a broad spectral range and probed mechanically by measuring the force gradient of the distance-dependent interaction force between the tip and the sample $F_{\text{ts}}$.~\cite{sifat_photo-induced_2022, Jahng2016, Wang_2019_generalizedHeterodyne, Springer2004} This enables the detection of the local chemical composition, molecular vibrations, and even chirality of the material to be imaged with high spatial resolution better than 10~nm.
\begin{figure}[t!]
    \centering    \includegraphics[width=\textwidth]{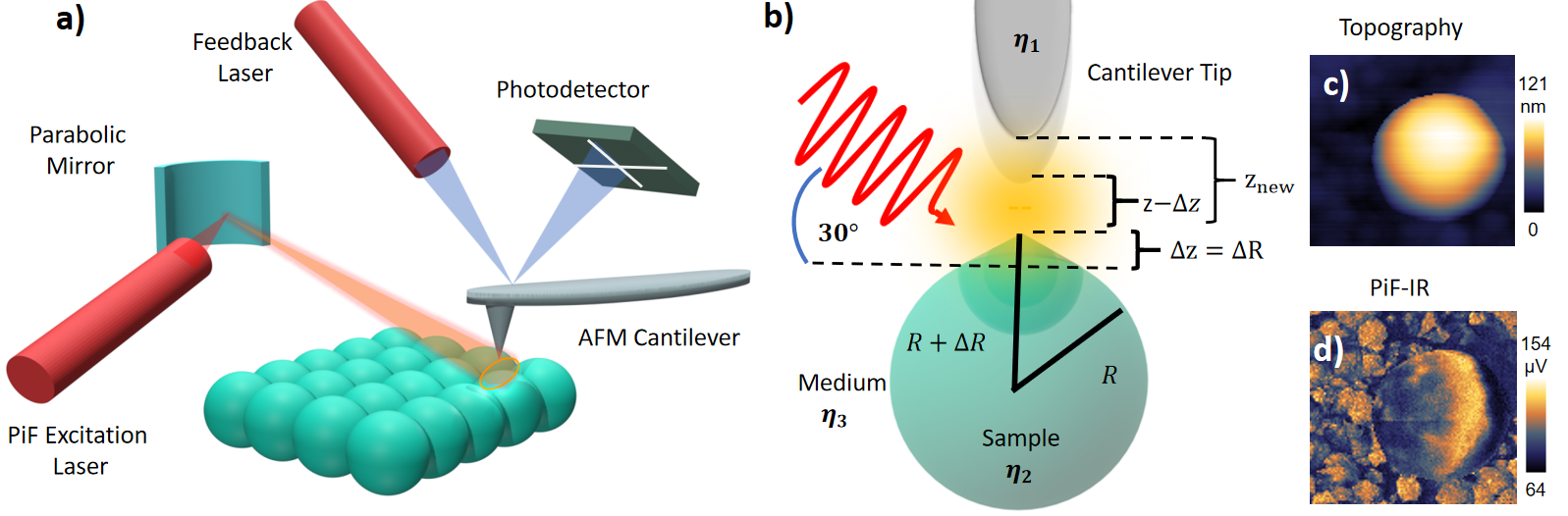}
    \caption{a) The basic PiFM diagram depicts how the AFM detects the PiF on a sample’s surface. b) Illustration of the effect of heat generation due to light absorption on the dynamic tip motion. Experimental results for c) topography, and d) PiF contrast of a spherical PMMA NP with a radius $R=$ 50~nm at 1150~cm$^{-1}$.}
    \label{fig1}
\end{figure}

As depicted in Fig.~\ref{fig1}a), light is focused onto the tip-sample region. The metal-coated AFM tip acts as an antenna for this light increasing the sample's absorptivity through tip-enhanced local fields.~\cite{Huth_antenna2013} The force acting between the sample and the tip changes as a result of the light absorption in the tip and the sample leading to photothermal expansion which modifies the tip oscillation. Previous studies have investigated the tip-sample dynamics on polymer slabs,~\cite{Jahng2022} heat generation by nanostructures~\cite{Quidant2013, Asgharian2020, Zograf:21} and NP arrays,~\cite{Baffou2013,Javier2010} as well as the influence of tip-artifacts and tip-deformations.~\cite{KOPYCINSKAMULLER2006466, Weisenhorn_1993, ricci_recognizing_2004} Several PiFM detection techniques, including sequential heterodyne, peak force tapping (PFT), homodyne, and heterodyne detection, have been developed and described theoretically.~\cite{Wang_2019_generalizedHeterodyne,Jahng2016, Wang2018} A recent study using a finite element method (FEM) for simulating an isolated AFM tip without any sample shows that the three-dimensional photo-induced force (PiF) can be affected by lateral photothermal vibration when the tip geometry is asymmetric.~\cite{Ritz2023_tipvibrations} 

Here we present a theoretical characterization of the photothermal expansion of nanostructured materials and its feedback to the tip-sample dynamics modifying the detected PiF. Our theoretical considerations are combined with the experimental evaluation of the PiF obtained from a spherical polymethyl methacrylate (PMMA) NP under varied mid-IR illumination conditions. In the theoretical part, we investigate several types of nanostructures relevant for PiFM and, in particular, for PiF-IR. We quantitatively analyze the effect of the sample's shape, the input laser power and the polarization on the thermal expansion. A detailed analysis of the force-distance behavior resulting from photothermal expansion is also presented. We discuss the effect of key parameters and study the static tip-sample interaction force $F_c$. We further simulate a PiF-IR scan over particles of different materials. The goal of our study is to provide a comprehensive, unified approach to PiF-IR on nano\-structured materials.


\section{Theoretical framework}
In the following, we give a short description of the theoretical framework, which is based on the considerations by Jahng et al..~\cite{Jahng2022, Jahng2019}  
Due to the near-field enhanced light absorption in the tip-sample region, the sample's surface expands by a distance $\Delta z$, see Fig~\ref{fig1}b). This change in the material density modifies the weak van der Waals (VdW) forces acting between the sample and the tip,
resulting in a change in the distance-dependent interaction force $F_{ts}$ between the sample and the tip,~\cite{Tauber2013, Jahng2019} which affects the dynamic tip motion. In PiFM, the cantilever scans in dynamic noncontact mode and a heterodyne detection scheme is employed to enhance the spatial resolution, i.e., the cantilever is driven at its second mechanical resonance frequency $f_{2}$ while the PiF is detected in side-band mode at the first mechanical resonance frequency $= f_{1}$ and the modulation frequency of the incident laser $f_{m}$ is set as either their sum or their difference: $f_{m} = |f_{2}\pm f_{1}|$. An example for the obtained PiF is given in Fig.~\ref{fig1}c) and d) showing the topography and the PiF contrast of a spherical PMMA nanoparticle (NP) with radius $R=50$~nm on a CaF$_2$ substrate, respectively. The diameter of the NP appears larger due the convolution of the NP geometry with the shape of the tilted cantilever tip.~\cite{ricci_recognizing_2004}
The PiF contrast was acquired using the mid-IR illumination frequency $\nu=1150$~ cm$^{-1}$, which is one of the absorption frequencies of PMMA. It reveals a distinctive asymmetry in the NP which results from the near-field coupling effects in the tip-sample junction.~\cite{Ritz2023_tipvibrations} According to Wang et al., the PiF is experimentally observable for up to 5~nm distance between tip and sample,~\cite{Wang_2019_generalizedHeterodyne} although, the antenna effect of metallic AFM tips is measurable for up to 20~nm.~\cite{Virmani2024-eo} The high surface sensitivity of PiFM is obtained by driving the cantilever oscillation at a small amplitude $A_2\approx$~1 to 2~nm,~\cite{Jahng2022} which is explained in more detail in the section on dynamic feedback.

We investigate the photo-induced expansion of NPs in this small oscillation limit using the gradient of $F_{\text{ts}}$. Starting from optical simulations of the absorption in the NP, we simulate the subsequent heat generation and surface expansion and calculate the resulting PiF using the heterodyne detection mode. By this, we demonstrate that different materials can be discriminated based on the measured force gradient.
 
\subsection{Static contribution to tip-sample interaction}
A general expression for the distance-dependent tip-sample force $F_{\text{ts}}(z)$ is given by~\cite{Jahng2022} 
\begin{equation}
    F_{\text{ts}}(z)=F_\text{c}(z)+F_{\text{nc}}(z)
\end{equation}
where $F_\text{c}(z)$ describes the contribution from conservative and $F_{\text{nc}}(z)$ non-conservative forces.
They are both functions of the tip-sample distance $z$. The conservative forces are given by
\begin{equation*}
    F_\text{c}(z) \approx 
    \begin{cases}
        \frac{-H_{\text{eff}}r}{12z^2}  & \text{for} \ (z>r_0) \\
        \frac{-H_{\text{eff}}r}{12 r_0^2}+\frac{4}{3}E^*\sqrt{(r_0-z)^3 r} \ \ \ \ &\text{for} \ (z\leq r_0)
    \end{cases}
\end{equation*}
Hereby, $H_{\text{eff}}$ is the effective Hamaker constant which describes the interaction energy between two bodies via a medium, $r$ is the tip radius and $E^*$ is the effective Young's or elastic modulus which is a measure of the stiffness and rigidity of the material. When the tip-sample distance is less than the interatomic distance $r_0$, the conservative force becomes the sum of two forces. The first one is the noncontact VdW force and the second term is the contact Derjaguin, Muller, and Toporov (DMT) force.~\cite{Jahng2022} The theoretical DMT model is used to estimate the adhesion force between two surfaces.
The Hamaker coefficient results from summing up the pairwise contributions of points opposite of each other of two large bodies interacting via a medium. This geometric approximation is valid, since cross-interactions approximately cancel due to the overlap of symmetric force contributions.~\cite{VANKAMPEN1968307,Israelachvili2011} For two bodies of different materials and shapes interacting in a medium it was evaluated by Lifshitz.~\cite{Israelachvili2011} According to Lifshitz' evaluation, the effective Hamaker coefficient $H_{\text{eff}}$ of a material 1 interacting with a material 3 via a medium 2 is given as 
\begin{multline}
    H_{\text{eff}}=\frac{3}{4}k_\text{B}T \left(\frac{\epsilon_1 - \epsilon_3}{\epsilon_1+\epsilon_3}\right)\left(\frac{\epsilon_2 - \epsilon_3}{\epsilon_2+\epsilon_3}\right)+ \frac{3h}{4\pi} 
     \int_{\omega_1}^{\omega_2} \left(\frac{\epsilon_1(i\omega) - \epsilon_3(i\omega)}{\epsilon_1(i\omega)+\epsilon_3(i\omega)}\right)\left(\frac{\epsilon_2(i\omega) - \epsilon_3(i\omega)}{\epsilon_2(i\omega)+\epsilon_3(i\omega)}\right) d\omega,
\end{multline}

where $k_\text{B}$ is the Boltzmann constant, $T$ denotes the temperature, $h$ is Planck's constant, the $\epsilon_\text{j}$ denote the static dielectric constant of material $\text{j}=1,2,3$ and the $\epsilon_\text{j}(i\omega)$ are the values of $\epsilon$ at imaginary angular frequencies $i\omega$ related to Matsubara frequencies evaluated~\cite{Esteso2024-vb} from 
\begin{align}
\epsilon\left(i\omega\right) = 1 + \frac{2}{\pi}\int_{0}^{\infty} \frac{\bar{\omega}\ \text{Im}(\epsilon(\bar{\omega}))}{\bar{\omega}^2 + \omega^2} d\bar{\omega}
\end{align}
with $\omega= \frac{2\pi c}{\lambda}$ is the angular frequency of the light for the vaccum wavelength $\lambda$. We use the angular frequency $\omega$ here to avoid confusion with the light frequency given in units of a wavenumber $\nu = 1/\lambda$, which is the common approach in mid-IR spectroscopy and is used in this work for presenting results.
\begin{figure}[t!]
    \centering
    \includegraphics[width=0.49\columnwidth]{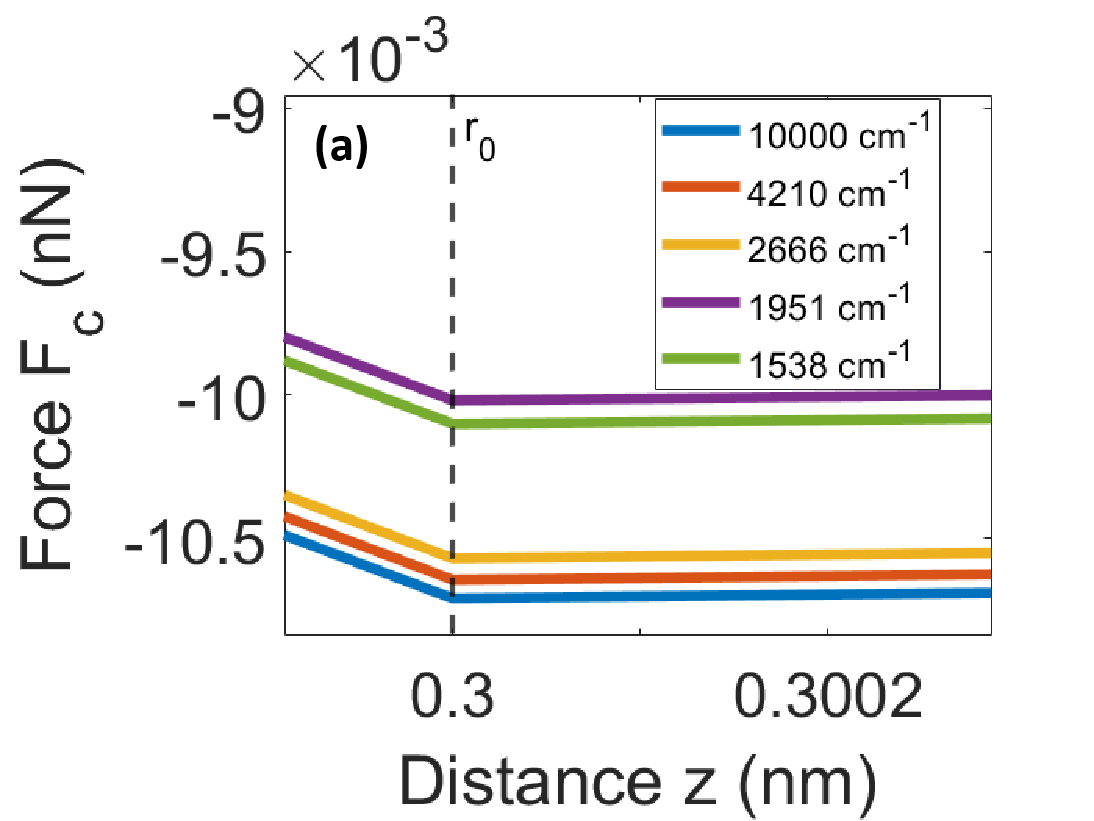}
    \includegraphics[width=0.49\columnwidth]{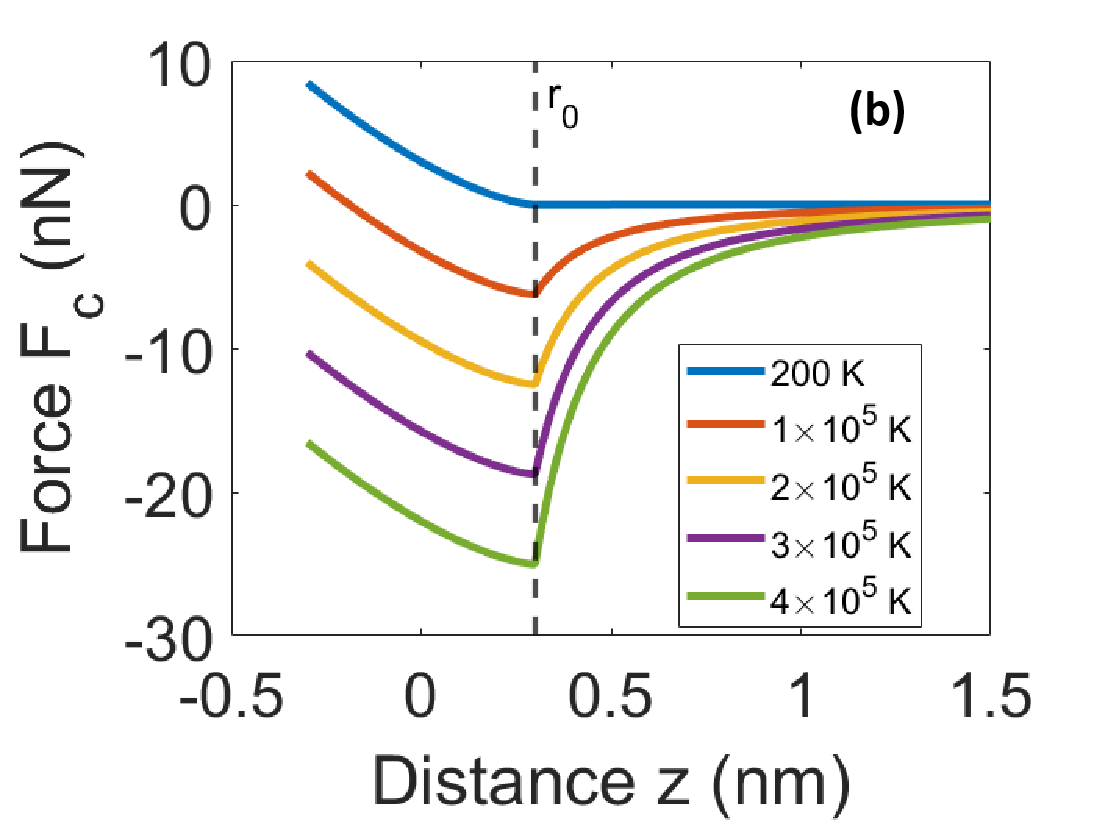}
    \includegraphics[width=0.49\columnwidth]{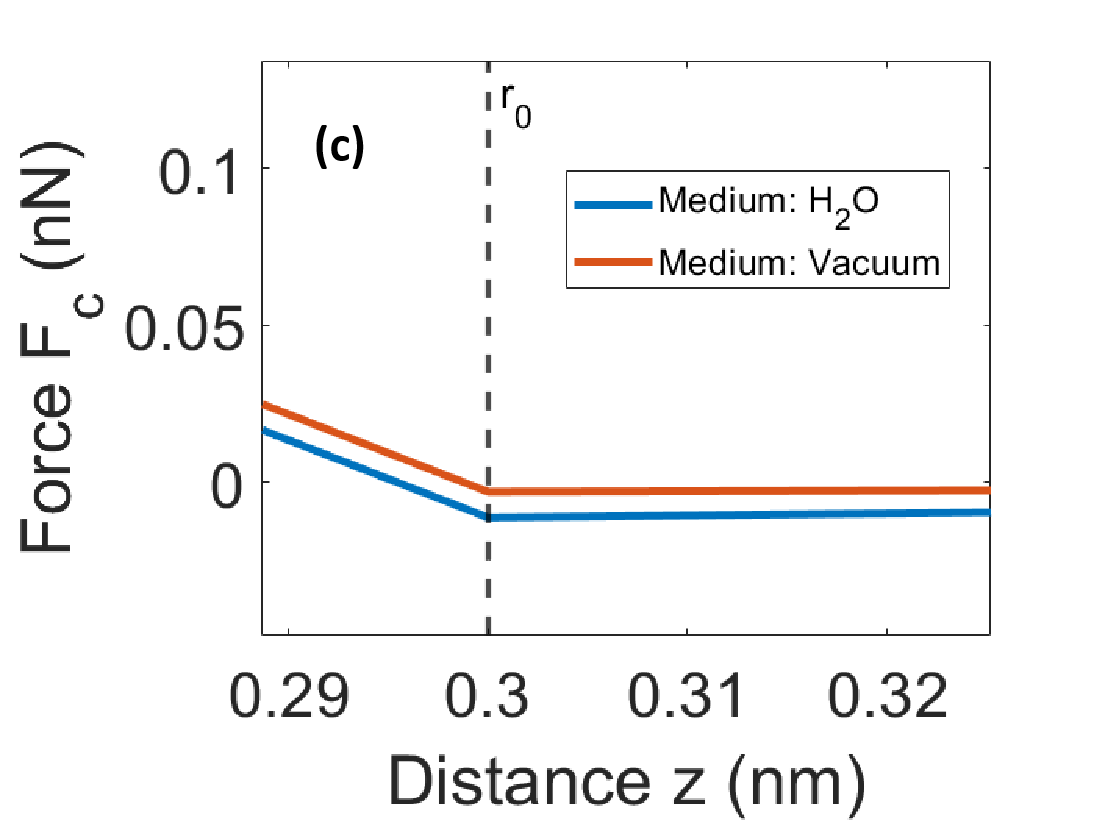}
    \includegraphics[width=0.49\columnwidth]{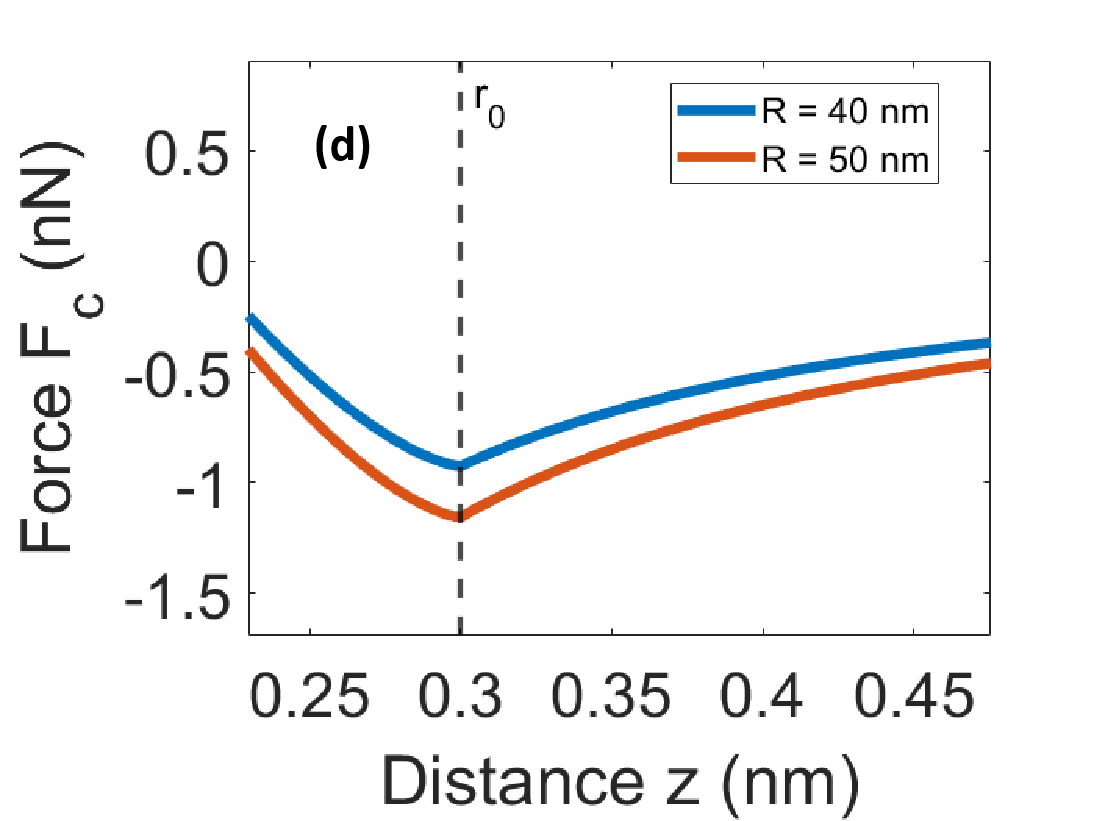}
    \caption{Static VdW forces for a Pt tip and a PMMA NP as a function of the tip-sample distance for varied (a) light frequency $\nu$, and for $nu=1730$~cm$^{-1}$ (b) temperature of the heated particle, (c) interaction medium, and (d) NP radius $R$.}
    \label{fig2}
\end{figure}

For our simulations, an interatomic distance of $r_0=0.3$~nm is considered, a typical value in AFM setups.~\cite{Jahng2022, Kabylda2023} The ultrasharp Pt tips used in our experiments typically are 10-15~$\mu$m long having a tip apex radius $r\approx 10$~nm. We model the tip accordingly as a 10~$\mu$m long Pt ellipsoid with a 180~nm diameter resulting in $r=10$~nm. We consider PMMA and polystyrene (PS) nanostructures such as slabs of 1~$\mu$m $\times$ 1~$\mu$m $\times$ 100~nm and spherical NPs with $R=40$ and $R=50$~nm as samples. In general, we use air as interaction medium, apart from one example given in Fig.\ref{fig2}, where water was used.

The angular frequency $\omega$ (or frequency $\nu$) of the illuminating light, the temperature, the material composition and the tip size contribute to $H_{\text{eff}}$ and thus impact the interacting force in the near contact region, see Fig.~\ref{fig2}. Even if tip and sample are of the same material, the $H_{\text{eff}}$ can vary depending on the surrounding medium. The frequency of the incident electric field can affect the $H_{\text{eff}}$ and therefore alter the interaction force, as depicted in Fig.~\ref{fig2}(a), even before considering light coupling effects and local field enhancement. While the change in $H_{\text{eff}}$ may not be readily apparent in realistic temperature ranges up to 400K, a significant change in temperature, see Fig.~\ref{fig2}b), can result in a stronger (negative) interaction force between the tip and sample. A comparison of the effect of the interaction medium on the force is given in Fig.~\ref{fig2}c) representing the tip-sample interaction in air and in water over distance $z$ for a Pt tip over a PMMA surface at wavelength $\lambda=480$~nm at room temperature. When the gap between tip and sample is larger than the interatomic distance, the force difference is relatively small. Conversely, when the gap becomes smaller than the interatomic distance, the force difference strongly increases since the contact DMT force acts together with the VdW force. Finally, the static force depends on the geometry of the sample, e.g. the nanometer scale particle radius $R$, see Fig.~\ref{fig2}d).

\subsection{From light absorption to thermal expansion}
The total electric field stems from contributions by direct illumination of the sample material as well as from field enhancement in the tip-sample junction due to optical coupling.~\cite{Jahng2019, Jahng2022}
The presence of the electric field $\vec {E}_{\text{NP}}$ created by incident light of angular frequency $\omega$ leads to heat generation inside the nanostructures.~\cite{Sallam:14, Zhou2016} 
The power absorbed by a NP is calculated from the divergence of the Poynting vector $\vec{S}(\vec{r})$ at each position $\vec{r}$
\begin{equation*}
    \nabla\cdot\vec{S}(\vec{r}) = \frac{1}{2}\epsilon_0 \omega \text{Im}(\epsilon)|\vec{E}_{\text{NP}}(\vec{r})|^2
\end{equation*}
which provides the absorbed energy per unit volume and relates to the properties of the medium via its absorption given by the imaginary part of its permittivity $\text{Im}(\epsilon)$ with $\epsilon_0$ being the free space permittivity. Calculating the total absorbed energy $Q_{\text{abs}}$ requires an integral over the volume of the relevant region, here, the absorbed electric field inside the nanoparticle.~\cite{Jahng2022} Including the effective thermalization time $\tau_{\text{th}}$ which depends on the pulse width $t_\text{p}$ of the excitation laser and the relaxation or cooling time $\tau_{\text{rel}}$, this results in the total absorbed energy
\begin{equation}
    Q_{\text{abs}}=\frac{1}{2}\tau_{\text{th}} \omega\epsilon_0 \text{Im}(\epsilon)\int |\vec {E}_{\text{NP}}|^2  dV_{\text{abs}}.
\end{equation}
$t_\text{p}$ contributes to $\tau_{\text{th}}$ through~\cite{Jahng2022}
\begin{equation}
    \tau_{\text{th}} = \tau_{\text{rel}}\left( 1 - e^{-\frac{t_\text{p}}{\tau_{\text{rel}}}} \right).
    \label{eq.thermalizationtime}
\end{equation}
The relaxation time $\tau_{\text{rel}}$ depends on the material
\begin{equation}
    \tau_{\text{rel}} = \frac{4}{\pi^2}\frac{\rho C}{\kappa_{\text{eff}}}l_z^2.
    \label{eq. relaxationtime}
\end{equation}
$\rho$, $C$, $\kappa_{\text{eff}}$ and $l_z=\sqrt{Dt_p}$ are the density of the sample material, total heat capacity, effective thermal conductivity, and heat diffusion length, and $D$ is the thermal diffusivity. For the materials used in our work, these parameters are listed in table S1 of the Supporting Information (SI).

From the distribution of the electric field, we deduce the absorption inside the nanostructure and calculate the generated heat. The heating power (energy over time) results in a local temperature change and is the cause of the observed thermal expansion. The heat diffusion equation can be used to determine the temperature distribution inside and around NPs.~\cite{Zograf:21,Baffou2010} Despite the fact that the effective thermal conductivity depends on geometry,~\cite{baffou2009heat} the tip-enhanced local field heats the entire NP, because the NP volume generally falls within the heat diffusion volume making the temperature distribution inside NPs nearly constant, however, in microsctructures it can be non-uniform. 

From thermodynamics, the resulting temperature change in the material $\Delta T$ can be deduced from the thermal energy
\begin{equation}
     Q_{\text{heat}} = c_p m\Delta T =c_p V\varrho\Delta T \equiv Q_{\text{abs}},
\end{equation}
where $c_p$, $m$, $V$, $\varrho$ are the specific heat capacity, mass and volume, and the mass density of the material, respectively. This assumes that the optical absorption is converted to heat energy without losses. The temperature rise causes the material to expand~\cite{Hummel1993} and we apply a linear thermal expansion to the NP of radius $R$ resulting in a change in $\Delta R$:
\begin{equation}
    \Delta R= R\alpha\Delta T = \Delta z,
\end{equation}
where $\alpha$ is the linear thermal expansion coefficient of the material.
In our simplified model, we assume the expansion to be directional in $z$-direction, i.~e., we consider only expansion which is parallel to the direction of the force $F_c$, see the sketch in Fig.~\ref{fig1}(b), therefore, no higher-dimensional thermal expansion coefficients are taken into account. Because of the larger illuminating wavelength $\lambda>R$ the NP expansion combines a component $\Delta R_d$ from direct illumination of the sample material and a component $\Delta R_t$ due to the tip-enhanced near-field, resulting in $\Delta R=\Delta R_d+\Delta R_t$.

\subsection{Dynamic feedback}
After the mechanical expansion of the surface, the difference in the tip-sample interaction force $F_{\text{ts}}$ becomes
\begin{equation}\label{eq.dynamic_feedback}
    \Delta F_{\text{ts}}(z) = F_{\text{ts}}(z-\Delta z) - F_{\text{ts}}(z)
\end{equation}
initiating a dynamic tip motion of the cantilever. 
A laser beam with a modulation frequency $f_m$ illuminates the tip-sample junction while the cantilever oscillation is driven at the frequency $f_2$ with amplitude $A_2$. Employing the heterodyne detection scheme, the instantaneous tip position can be expressed as $z(t)\approx z_0 + z_2(t) + z_s(t) - \Delta R$. Hereby, $z_0$ is the average tip-sample distance, $z_2(t)$ is the coordinate of the driven oscillation at the second mechanical resonance frequency and $z_s(t)$ that of the heterodyne motion at $f_s = |f_m \pm f_2|$. Since the tip oscillates close to the surface, a frequency-dependent force spectrum is produced based on the change of the tip-sample interaction force $F_{\text{ts}}$, which is caused by the thermal expansion. The dynamic tip motion becomes~\cite{Jahng2022} 
\begin{equation}\label{eq.extended_tipsample_force}
    F_{\text{ts}} \approx  F_c(z_0) + \frac{\delta F_c}{\delta z} (z_2) - \Gamma\dot z_2 +F_s,
\end{equation}
where $\Gamma$ is the damping of the cantilever movement and $F_s$ is the heterodyne PiF expressed as
\begin{multline}\label{eq. Fs}
    F_s \approx -\frac{1}{2}\left(\frac{\delta F_c}{\delta z} \frac{\delta R_t}{\delta z}|_{z_0} + \frac{\delta^2 F_c}{\delta z^2} \Delta R \right) A_2 \sin(f_st+\theta_s) -\frac{1}{2}\Gamma\omega_s\frac{\delta R_t}{\delta z}|_{z_0}A_2\cos(f_st+\theta_s),
\end{multline}
where $\theta_s$ is phase of the heterodyne motion. Depending on the operation mode, different terms dominate the measured PiF. The homodyne PiF is directly proportional to the thermal expansion,~\cite{Jahng2022} but the here employed heterodyne PiF ($=F_s$) is sensitive to the tip-enhanced thermal expansion $R_t$ in the small oscillation limit, where the first derivative term dominates the second one in Eq.~\ref{eq. Fs} resulting in:~\cite{Jahng2022,Jahng2019}
\begin{equation}\label{eq.thermal force}
    F_s \approx \frac{\delta F_{\text{ts}}}{\delta z}\Delta z.
\end{equation}
Thus, in the heterodyne detection mode, the tip-enhanced thermal expansion is dominated by the noncontact interatomic tip-sample force, because the signal vanishes in the contact region as well as far from the sample, which results in a high surface sensitivity.~\cite{Jahng2019}

\section{Methods and Materials}
\subsection{Simulation}
The Boundary Element Method (BEM),~\cite{Hohenester2019} in particular, the MATLAB-based simulation toolbox MNPBEM (Metallic Nanoparticles (MNPs) BEM,~\cite{toolbox1,toolboxeels,toolboxsubstrate,HOHENESTER2018209} is used to compute the electric field forming around the structure from induced surface charges. The toolbox equally applies to dielectric and hybrid setups as for the tip-sample interaction considered here. The Supporting Information (SI) includes a brief overview of the simulation procedure and discusses the importance of retardation effects. For this, Fig. S1 shows the temperature change described below comparing the results for the quasi-static and fully retarded calculation. Moreover, details on the surface discretization used in BEM calculations are given. The dimensions of either the polymer slab or the spherical NP are 1\ $\mu$m $\times$ 1\ $\mu$m $\times$ 100~nm, and  $25\le R\le 50$~nm, respectively. We employ a 10 $\mu$m long ellipsoidal Pt tip with either a 180~nm radius achieving $r=10$~nm or 230~nm for $r=20$~nm, resp., to ensure compatibility with the experimental results. We compare the PiF obtained for 60$\degree$ and for 80$\degree$ incidence using an illumination duration of 10~ns. Apart from one example, where the medium for the interaction between tip and sample is water, all calculations are done with air as medium, for which we use the vacuum frequencies of light.
\textcolor{blue}{The simulation code is published on github.}

\subsection{Experimental}
PiF-IR measurements were conducted using a VistaScope$^{\rm {TM}}$ (Molecular Vista Inc., US) equipped with a tunable and pulsed quantum cascade laser (QCL, Block Engineering, US) which illuminates the tip-sample region at a high angle of incidence ($\approx 80\degree$ with a focal spot diameter in the range of twice the illumination wavelength ($2\lambda$). A high-frequency noncontact ultrasharp PtIr-coated PointProbe$^{\rm {TM}}$ Plus (PPP) cantilever (Nanosensors$^{\rm {TM}}$, CH) was operated mainly in the attractive regime of the tip-sample interaction force by choosing a set-point of 85\% of the $\approx 2$~nm amplitude of the driven oscillation. Pt-coated PPP cantilevers typically have a 42 ${\rm {Nm}}^{-1}$ force constant and a first mechanical resonance frequency $f_1=$ 330 kHz. Typical feature of PPP tips are: tip apex radius $r\le 7$~nm, tip height = 10-15 $\mu$m, half cone angle at the tip apex = 10$\degree$. PiF-IR contrast images and the PiF-IR hyperspectrum were acquired in heterodyne (sideband) detection mode, driving the cantilever at its second mechanical resonance frequency $f_2$ and measuring the PiF at the system’s first mechanical resonance frequency $f_s = f_2 - f_m = f_1 \approx 300$ kHz, while the height image (AFM topography) was recorded at the driving frequency $f_2 \approx$1650 kHz. The QCL was pulsed at the frequency $f_m = f_2 - f_1$ to enhance the detected signal.~\cite{sifat_photo-induced_2022, joseph_nanoscale_2024}
The PiF-IR hyperspectrum was acquired using a homogenized illumination power of about 100 $\mu$W by clipping the previously recorded power spectrum at 5\% of its peak intensity and smoothed by a Savitzky-Golay filter 2-9. 
PiF contrast images at 1150 and 1300 ${\rm {cm}}^{-1}$ illumination frequencies were acquired at varied illumination power in the range of a few 100 $\mu$W. Scan images were processed using SurfaceWorks$^{\rm{TM}}$ 3.0 Release 30 (Molecular Vista Inc, US).

An aqueous suspension of PMMA nanospheres with diameter $100\pm 5$~nm (microParticles GmbH, Germany) was further diluted and then dripped on a calciumfluoride window (Crystal GmbH, Germany) and dried in air.

\section{Results and Discussion}
\subsection{Thermal expansion of a PMMA slab in resonant condition}
\begin{figure}[t!]
    \centering
    \includegraphics[width=0.49\columnwidth]{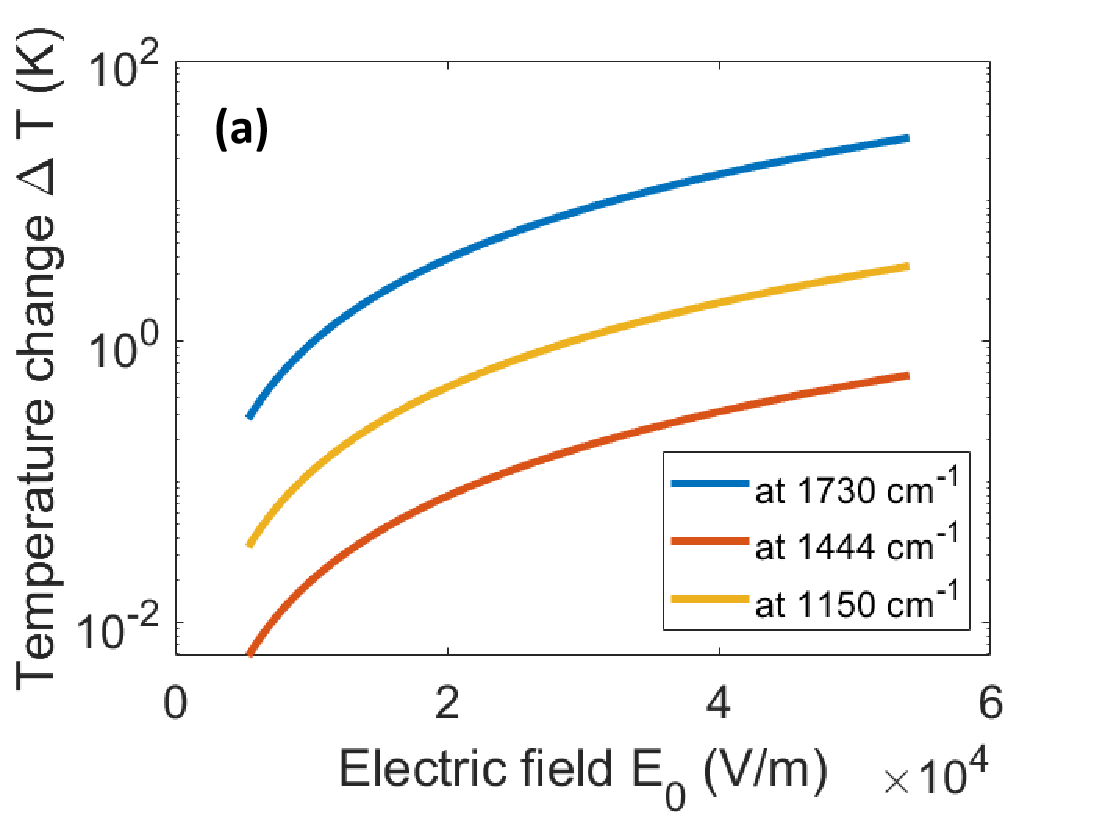}
    \includegraphics[width=0.49\columnwidth]{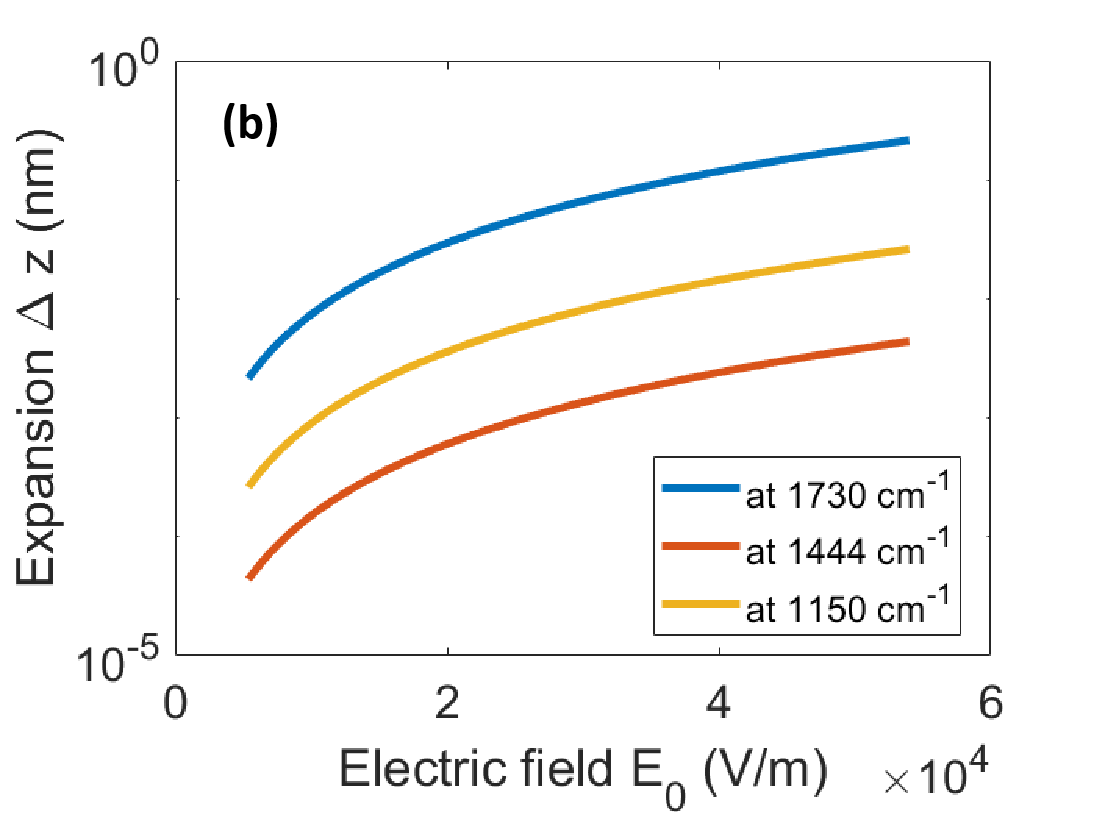}
    \caption{a) Temperature change and b) expansion of a PMMA slab (1 $\mu$m $\times$ 1 $\mu$m $\times$ 100~nm) under 80$\degree$ light incidence at three different $\nu$ matching absorption frequencies of PMMA.}
    \label{fig3}
\end{figure}
Before discussing the effect of the photo-induced thermal expansion of nanostructures on the obtained PiF, we model the temperature change and associated surface expansion of a PMMA slab of dimension 1 $\mu$m $\times$ 1 $\mu$m $\times$ 100~nm probed by a Pt tip at three different $\nu$ matching absorption frequencies of PMMA, see Fig.~\ref{fig3}. The angle of the incident light is kept at 80$\degree$ in accordance to realistic experimental setups.  The chosen frequencies $\nu$ = 1150~cm$^{-1}$ (6896~nm), $\nu$ = 1444~cm$^{-1}$ (6925~nm), and $\nu$ = 1730~cm$^{-1}$ (5780~nm) match specific bond vibrations in PMMA. The strongest of these is the C=O stretching vibration, which peaks at 1730~cm$^{-1}$. At this frequency, a temperature change of up to 27.8K is achieved for an incident field amplitude up to 54 $\times$ 10$^3$ V/m. 
The absorption band at $\nu$ = 1150~cm$^{-1}$ includes the C-O-C stretching vibration. The CH$_2$ bending vibration at $\nu$ = 1444~cm$^{-1}$ involves less change in the dipole moment. This reduces the temperature change and expansion at $\nu$ = 1444~cm$^{-1}$ compared to the other two absorption frequencies, see Fig.~\ref{fig3}. The obtained values agree with expectations and with published results from other groups.~\cite{Jahng2022}

\subsection{Effect of light polarization}
For the investigation of nanostructures, we have to consider the effect of both vertical and lateral displacement of the tip under varying illumination conditions. When plasmonic particles are brought into proximity, their individual plasmon resonances can couple. The displacement of the tip results, hence, in a complex field enhancement pattern based on asymmetric tip-particle plasmon coupling yielding a collective response known as tip-induced plasmon or gap plasmon that can significantly enhance electromagnetic fields in the surrounding medium.~\cite{bookUlrich,Maier2007-sj} Similarily, optical coupling takes place in hybrid systems where, e.g., a Pt tip interacts with a polymer surface. This coupling is maximized where the individual particle's fields overlap spatially and spectrally. For sake of clarity, we provide results on both a resonant plasmonic system (Au NP - Pt tip) and a hybrid system (PMMA NP - Pt tip) under varying illumination conditions. 

\begin{figure}[t!]
    \centering
    \includegraphics[width=1\columnwidth]{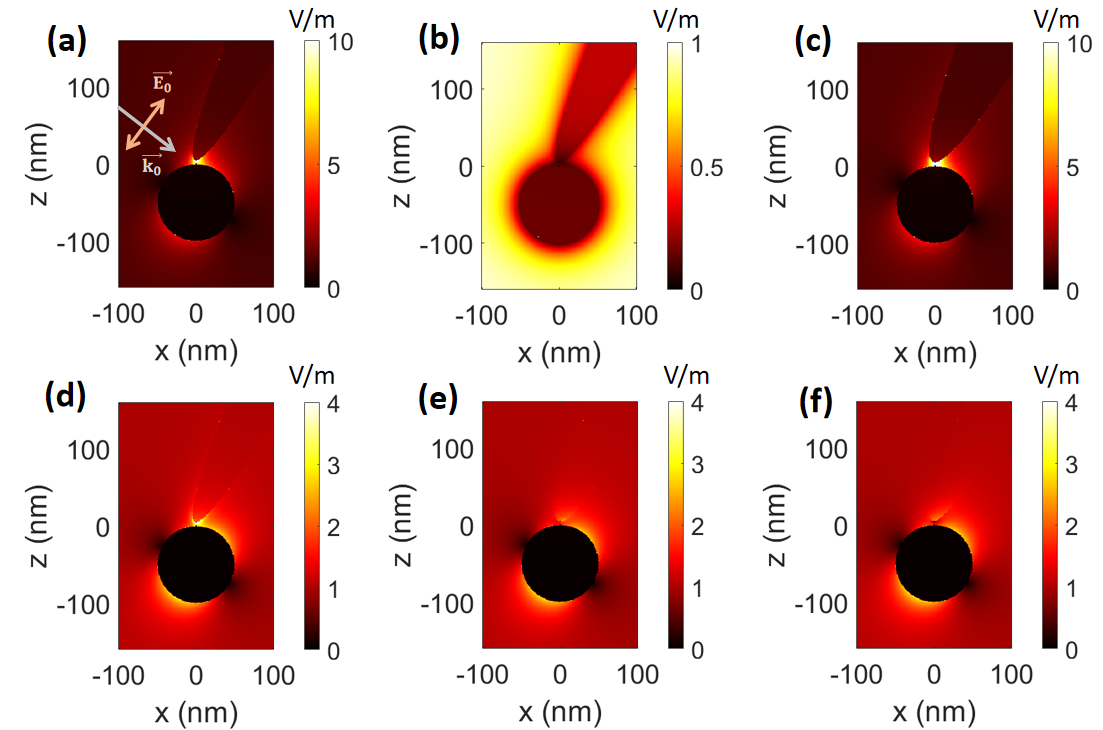}
    \caption{Field distribution in $x$--$z$ of spherical NPs under an Pt tip inclined at $30\degree$ for (a-c) a Au NP with $R=50$~nm with (a) $r=10$~nm tip apex and P-polarization, (b) S-polarization, (c) $r=20$~nm and P-polarization, (d - f) a PMMA NP illuminated under P-polarization with $R=50$~nm and $r=10$~nm using (d) $\nu=1730$~cm$^{-1}$ and (e) $\nu=1150$~cm$^{-1}$ within PMMA absorption bands, and (f) $\nu=1300$~cm$^{-1}$ outside those bands.}
    \label{fig4}
\end{figure}

The polarization direction of the light illuminating the tip-sample junction under a given angle of incidence is crucial for the coupling strength and, thus, the field enhancement. Therefore, light polarization plays an important role for achieving high surface sensitivity. For an Au NP under P-polarization, the gap plasmon can form due to the electric fields being in-plane with the particles and the gap between them allowing for plasmon coupling, as can be seen in the heatmaps in Figs.~\ref{fig4}(a) and (c) presenting the normalized field strength in the $x$-$z$-plane. On the other hand, as shown in Fig.~\ref{fig4}(b), for S-polarization the fields oscillate out-of-plane hindering significant spatial overlap of the plasmonic fields. Hence, we concentrate for our subsequent results on P-polarization only. For completeness, we consider S-Polarization for a polymer NP as a function of the lateral and vertical displacement of a Pt tip in Fig. S2. When a tip with a larger apex radius $r$ is used under P-polarization, the coupling effect is enhanced, see Fig.~\ref{fig4}(c). The increased local field outside the NP, see Figs.~\ref{fig4}(a) and (b), reduces the energy absorbed by the NP due to the overall energy conservation which holds true for changing the tip apex radius. 

Figs.~\ref{fig4}(d-f) explore this setup for PMMA nanospheres with $R=50$~nm under P-Polarization and $r=10$~nm. Here, we vary the excitation frequency from $\nu=1730$~cm$^{-1}$ and $\nu=1150$~cm$^{-1}$ within PMMA absorption bands (resonant conditions) in Fig.~\ref{fig4}(d,e) to $\nu=1300$~cm$^{-1}$ beyond such bands in Fig.~\ref{fig4}(f) as an off-resonant case. Of course, the illumination frequency also affects the formation of the tip plasmon. As can be seen from our simulation in Figs.~\ref{fig4}(d-f), there is a strong plasmonic field at the tip apex for our modeled tip. Such fields have been observed also experimentally for metallic tips.~\cite{Huth_antenna2013} Overall, the observations for the hybrid system are similar to an equivalent plasmonic system, however, the field enhancement due to optical coupling is weaker. Nevertheless, there is a considerable field coupling to induce a tip-enhanced thermal expansion.

\begin{figure}[t!]
    \centering
    \includegraphics[width=0.49\columnwidth]{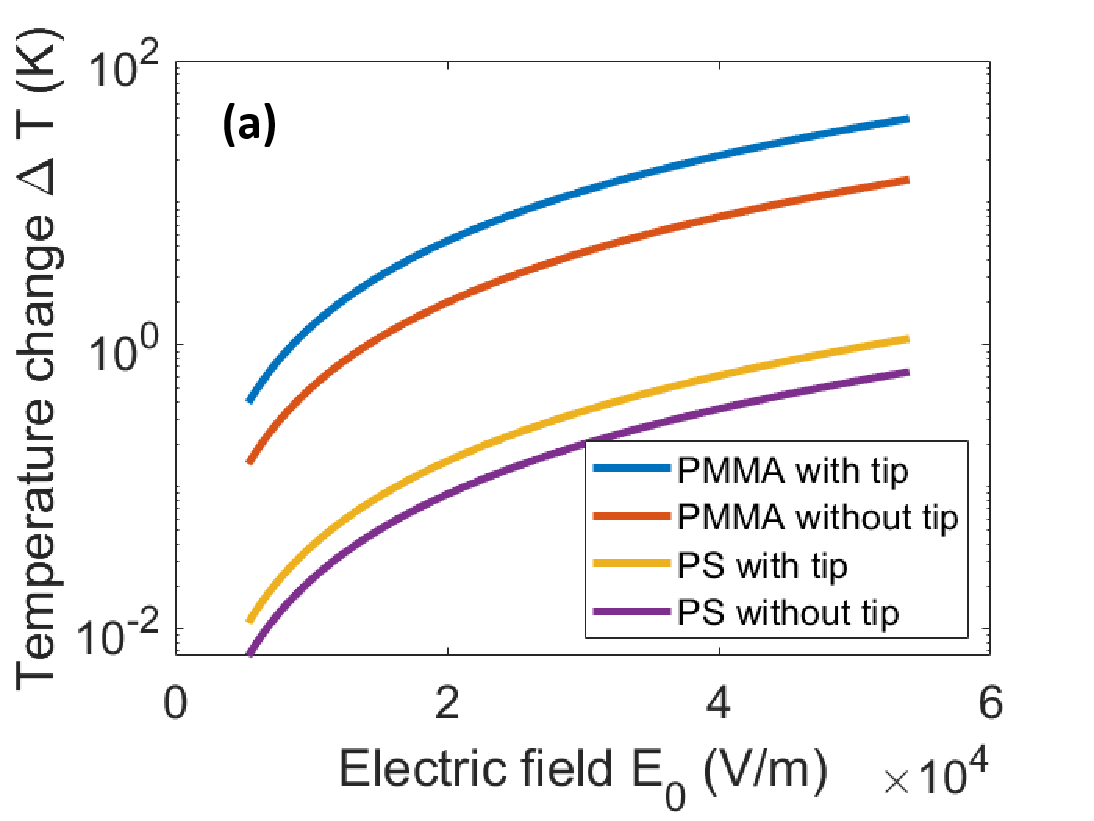}
    \includegraphics[width=0.49\columnwidth]{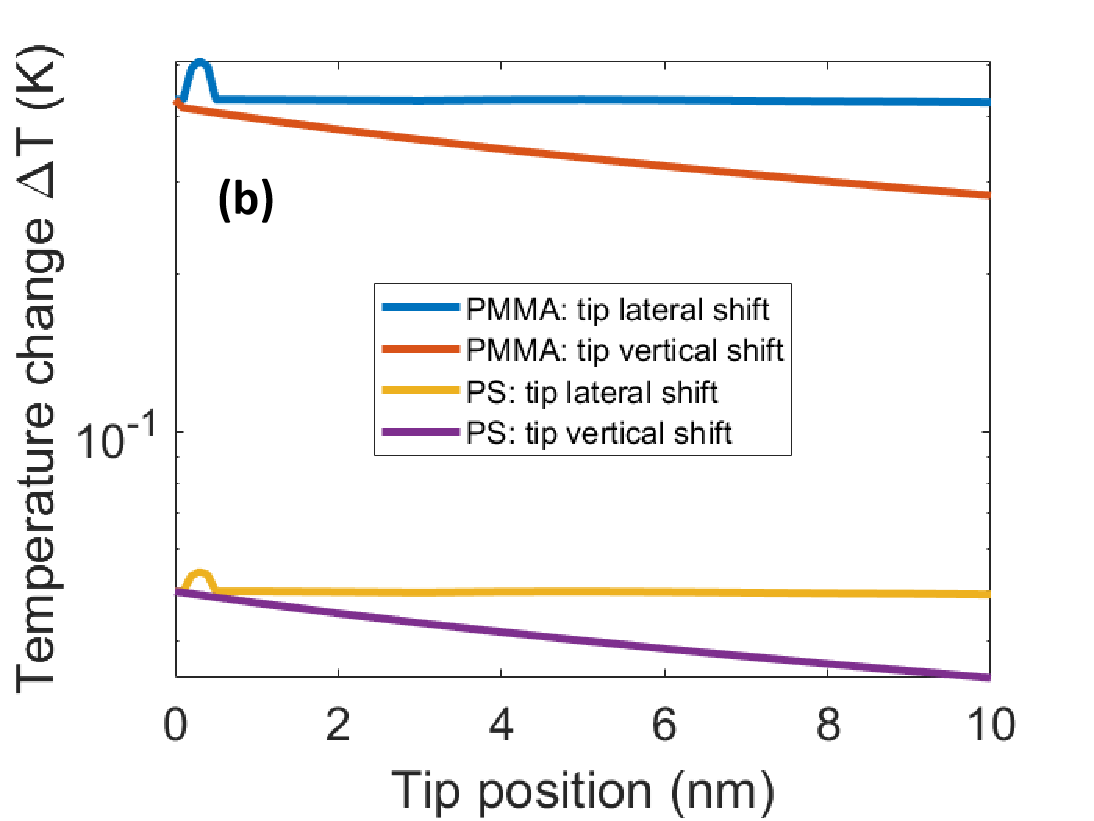}
    \includegraphics[width=0.49\columnwidth]{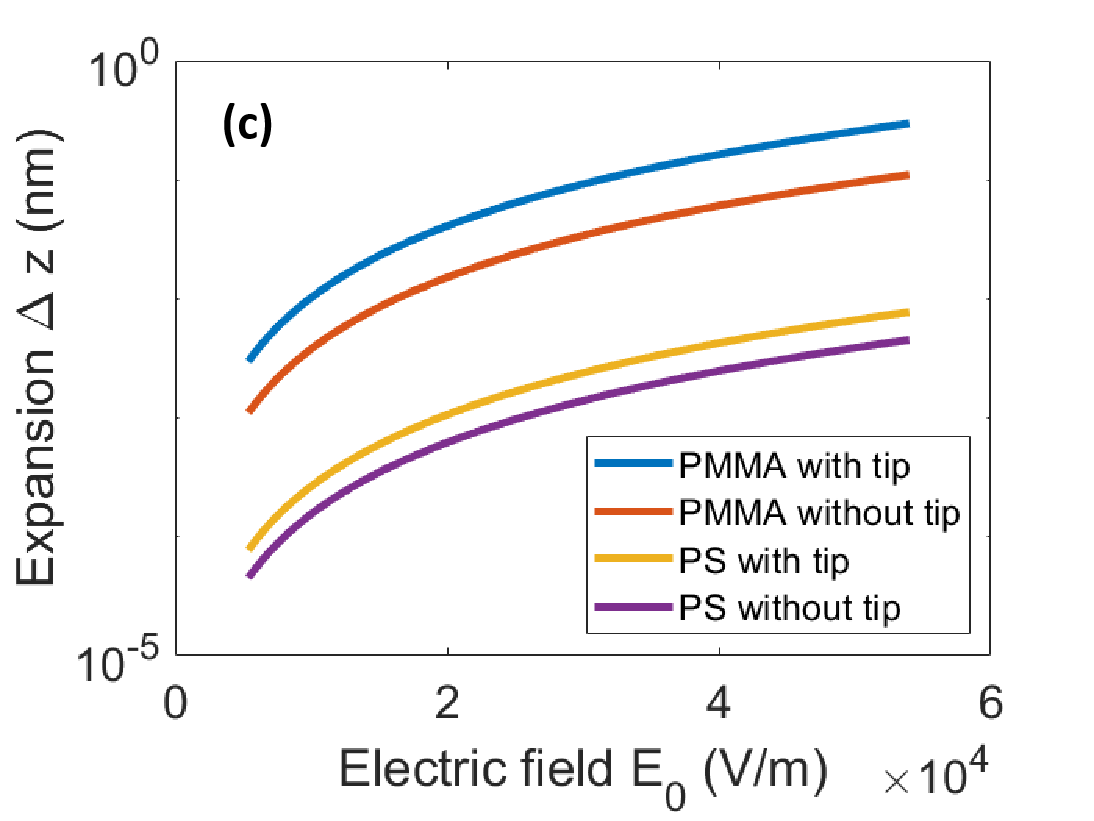}
    \includegraphics[width=0.49\columnwidth]{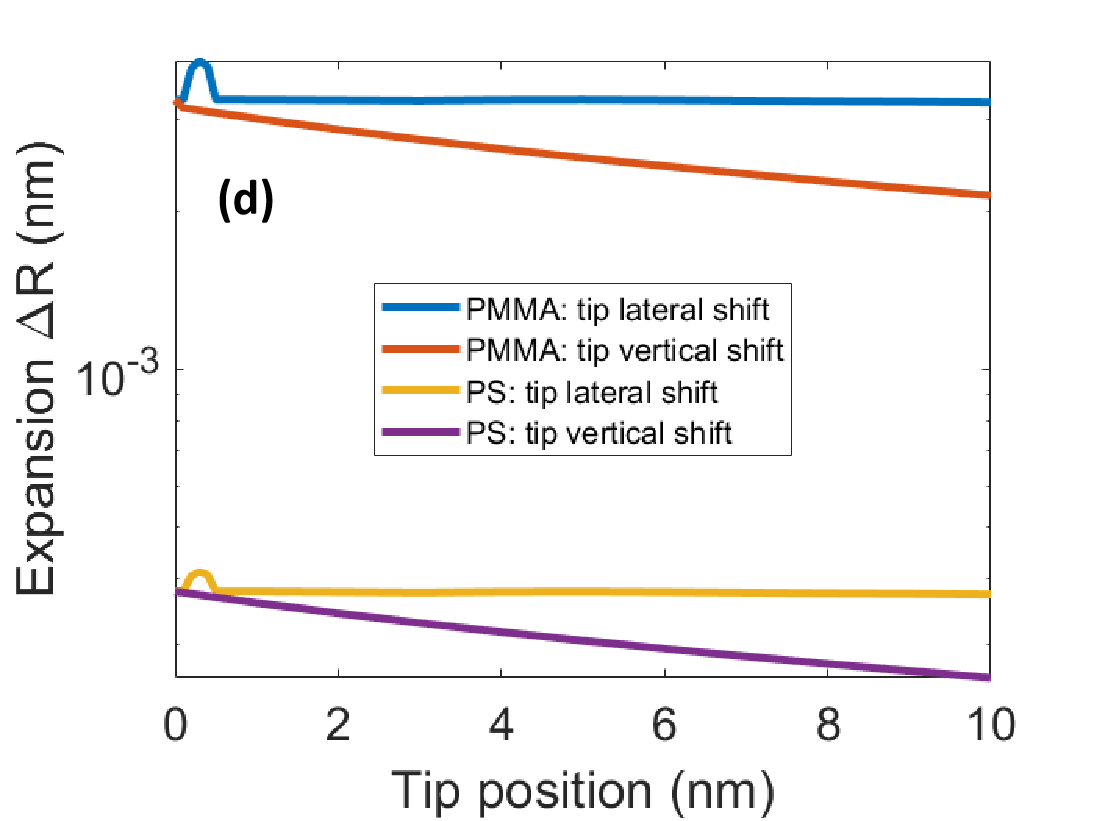}
    \caption{Temperature change $\Delta T$ at 60$\degree$ of light incidence under resonant conditions for (a) a polymer slab as a function of the incident laser power, and (b) a spherical polymer NP with $R=50$~nm for different lateral and vertical tip positions and associated expansion (c) $\Delta z$ of the slab, and (d) $\Delta R$ of the NP for either PMMA at $\nu=1730$~cm$^{-1}$ or PS at $\nu=1492$~cm$^{-1}$ matching the respective absorption bands.}
    \label{fig5}
\end{figure}

Heat generation and thermal expansion of a NP both maintain a linear proportional relationship with the absorbed electric field. As a result, varying tip positions lead to a minimal change in temperature and surface expansion when S-polarization is used. The resulting expansion may not be crucial for detection of the topology as PiFM detects the distance-dependent force gradient, however, the material composition might no longer be detectable. 

Since there is a significant tip-enhanced field in P-polarization, the related thermal expansion dominates compared to S-polarization where only the bulk thermal expansion takes place at weaker local fields. This results in S-polarization being volume sensitive while P-polarization is being surface sensitive and better suited to detect not only surface morphology, but due to the different coupling strength and resonance frequencies of varying materials also material properties. These results from our simulation are in accordance with reports in literature for polymer slabs.~\cite{Jahng2022} Therefore, for the remainder of this article, we consider only P-polarization aiming to identify distinct NPs based on the force observed by the tip which traces back to the material itself.

\subsection{Heat generation and thermal expansion}
The thermal energy that the polymer slab or spherical NP produce due to interaction with the incident laser light is strongly affected by the tip's presence. Here we consider an incident angle of 60$\degree$ and the tip to be positioned 1~nm above the polymer slab. For PS and PMMA, we compute the local temperature change and associated expansion under resonant conditions (for absorbing frequencies) at 1492~cm$^{-1}$ (6702~nm) and 1730~cm$^{-1}$ (5780~nm), respectively, where absorption coefficients are known.~\cite{Jahng2022} The illumination field amplitude is increased up to 54 $\times$ 10$^3$ V$/$m in Fig.~\ref{fig5}(a), which is sufficiently low to be non-invasive. At $E_0=54\times10^3$~V/m, a temperature change of up to $\Delta T=14.5$~K can be achieved for PMMA and $\Delta T=0.6$~K for PS when no tip is present. After including the Pt tip, the temperature change increases significantly up to $\Delta T=39.3$~K for PMMA and $\Delta T=1.1$~K for PS, which is approximately 11.5~K higher compared to the case of light being incident at an angle of 80$\degree$ with an expansion of $\Delta z=0.22$~nm, see Fig.~\ref{fig3}.

As a result of the temperature change, the surface expands. With inclusion of the tip and the resulting plasmon coupling, a physical expansion can be achieved of up to 
$\Delta z=0.29$~nm for PMMA slab and $\Delta z=0.01$~nm for PS slab, Fig~\ref{fig5}(c). The material characteristics determine the thermal expansion of individual materials. Although the linear thermal expansion coefficients $\alpha=70\times10^{6}$~K$^{-1}$ of PS and $\alpha=76\times 10^{6}$~K$^{-1}$ of PMMA are comparable, the absorption coefficient of PMMA (8045.61~cm$^{-1}$) is one order of magnitude higher than that of PS (708.47~cm$^{-1}$) at the selected vibrational resonances.~\cite{Jahng2022} Table S1 in the SI lists the properties of the materials that are used to determine heat generation, such as the heat capacity, thermal conductivity, and thermal diffusivity.

\begin{figure}[t!]
    \centering
    \includegraphics[width=0.64\textwidth]{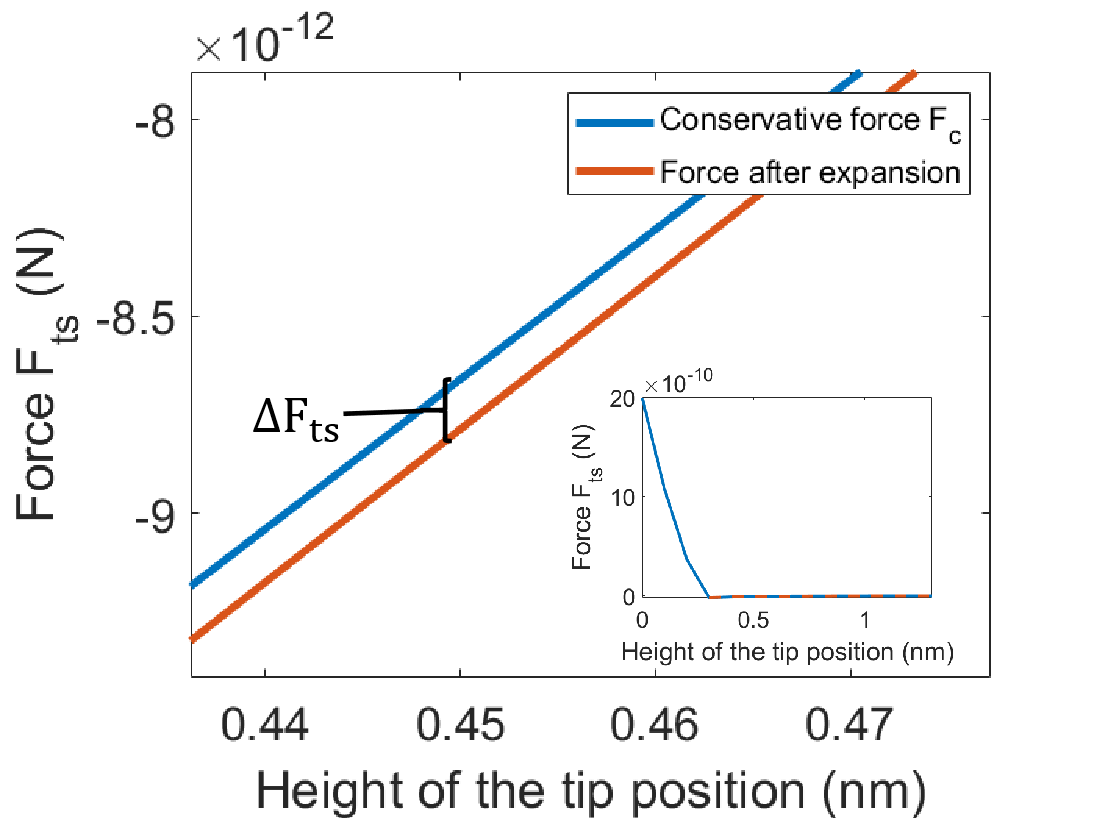}
    \caption{Force before and after thermal expansion of a spherical PMMA NP with $R=50$~nm excited at $\nu=1730$~cm$^{-1}$ in the absorption band of PMMA. The electric field amplitude is set to 54 $\times$ 10$^3$ V/m.}
    \label{fig6}
\end{figure}

\begin{figure}[t!]
    \centering
    \includegraphics[width=0.64\textwidth]{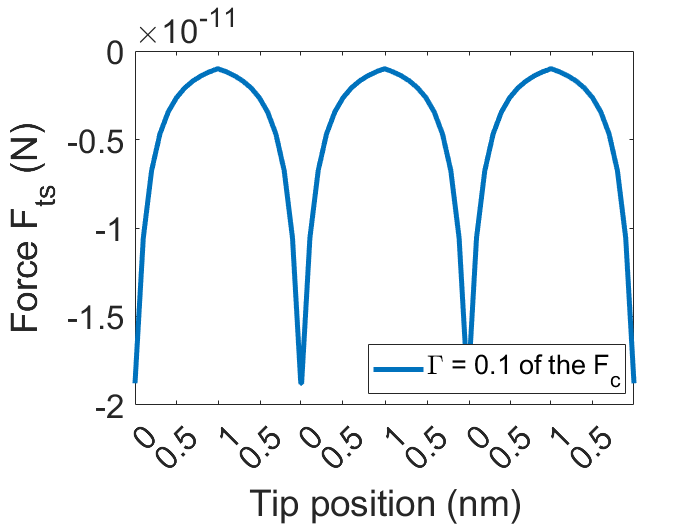}
    \caption{Dynamic feedback to the tip motion for the photothermal expansion of a spherical PMMA NP with $R=50$~nm: instantaneous tip-sample force F$_{ts}$ for a driven oscillation at $f_2= 1.8$~MHz with Amplitude $A_2=1$~nm and laser modulation at $f_s=2.1$~MHz.
    }
    \label{fig7}
\end{figure}

Figs.~\ref{fig5}(b) and (d) show that during lateral displacement of the tip, the temperature change $\Delta T$ and the surface expansion $\Delta z$ for the slab and $\Delta R$ for the NP become constant after a strong fluctuation when the surface of the NPs is close to the tip. This significant increase is due to the plasmon coupling effect, which results in $\Delta T=0.5$~K at a lateral distance of $\Delta x=0.3$~nm from the PMMA NP's vertex and in $\Delta T=0.1$~K for PS in that case. Consequently, the PMMA and PS NPs exhibit surface expansions of 
$\Delta R=3.9$~pm and $\Delta R=0.38$~pm, respectively, at that tip position as shown in Fig.~\ref{fig5}(d). With the vertical displacement of the tip a gradual decay is observed in both temperature change and expansion. In our simulation, both, the PMMA and the PS NP have $R=50$~nm. The expansion further affects the tip-sample interaction force F$_{ts}$ (see Fig. \ref{fig7}), causing a change in oscillation which we discuss quantitatively in the next section. 

\subsection{Dynamic feedback on multiple NPs}
First, we consider a spherical Pt tip with $r=10$~nm probing vertically over a spherical PMMA NP with $R=50$~nm as shown in Fig.~\ref{fig1}(b). Initially, the tip is oscillating over the sample's surface, where the minimum distance is assumed as $r_0=0.3$~nm, which is the position of contact with the NP's top surface. The amplitude of the oscillation of the AFM tip is considered to be $A_2=1$~nm, because beyond this distance, no significant change in the distance-dependent VdW force is observed~\cite{Jahng2022}. The conservative force $F_c$ acts between the tip and the NP. 

We now illuminate the tip-sample junction at $\nu=1730$~cm$^{-1}$ (5780~nm), which is one of the absorption frequencies of PMMA. The NP expands due to the abovementioned local heating of the surface. Due to this expansion, the distance between the tip and the sample is reduced, as can be seen from Fig.~\ref{fig5}(d). Therefore, the tip experiences a modified force $\Delta F_{\text{ts}}(z)$ given by Eq.~\ref{eq.dynamic_feedback}. During its oscillation it will, for example, at a new tip-sample distance $z'=0.45$~nm experience the increase in $\Delta F_{\text{ts}}(z'=0.45$~nm$) = F_{\text{ts}}(z'-\Delta z) - F_{\text{ts}}(z')\approx 0.13$~pN, see Fig.~\ref{fig6}, whereas,
$\Delta F_{\text{ts}}(z'=1.3$~nm$)\approx 5$~fN, which is no longer detectable.

From  we see that  due to the induced photothermal expansion while at a tip-sample distance of 1.3~nm the force change is reduced to around 5~fN which is no longer detectable. The minimum detectable force in PiFM can be calculated from  $F_{\text{min}}=\sqrt{4k_BTBk_i/Q_{f_i}f_i}$, where $f_i$, $Q_{f_i}$ and $k_i$ are frequency, quality factor and stiffness of the cantilever at its $i$-th mechanical resonance frequencies, respectively. $B$ is the bandwidth and $T$ the absolute temperature. A PPP-NCHAu cantilever has stiffness k$_1$ = 30 N/m and k$_2$ =1180 N/m and quality factor $Q_{f_1}$ = 360 and $Q_{f_2}$= 483 resulting in a minimum detectable force of 0.08 pN and 0.18 pN for the first two mechanical resonance frequencies, respectively.~\cite{Jahng2022, Jahng2016} We employ a Pt tip in our simulation matching the situation in the experiment.  The PPP-NCHPt tips typically have a quality factor $Q_{f_1} = 11860$~N/m~\cite{10.1063/1.4975629} and stiffness k$_1 = 42$~N/m which provides 0.02~pN as the minimum detectable force.      

After surface expansion, the tip experiences the distance-dependent conservative force $F_{c}$ and an additional (non-conservative) damping force $F_{nc}\sim\Gamma \dot z$, see Eq.~\eqref{eq.extended_tipsample_force}, which additionnally to material properties contributes to the total damping $\Gamma$. The resulting dynamic $F_{\text{ts}}=F_{c}+F_{nc}$ can be obtained at a given time or tip position. These mechanical driving forces are crucial for the sensitivity and surface response in PiFM. As detailed in the theoretical framework, we consider a heterodyne detection scheme employing $F_{s}$, for which we set $f_s=f_2+f_m = 2.1$~MHz. We consider a vertical deflection without lateral motion of the tip and a single 10~ns pulse starting at the surface of the PMMA NP and use a mechanical damping $\Gamma$ of 10\% of the conservative force $F_c$. The force gradient and the interactions that the tip experiences are also responsible for the occurrence of nonlinearity in PiFM.~\cite{Jahng2023} However, in the limit of small oscillation amplitudes, nonlinear effects can be ignored yielding a linear harmonic oscillation between the tip and the sample.~\cite{Jahng2016, Jahng2019}       
\begin{figure}[t!]
    \centering
    \includegraphics[width=0.49\columnwidth]{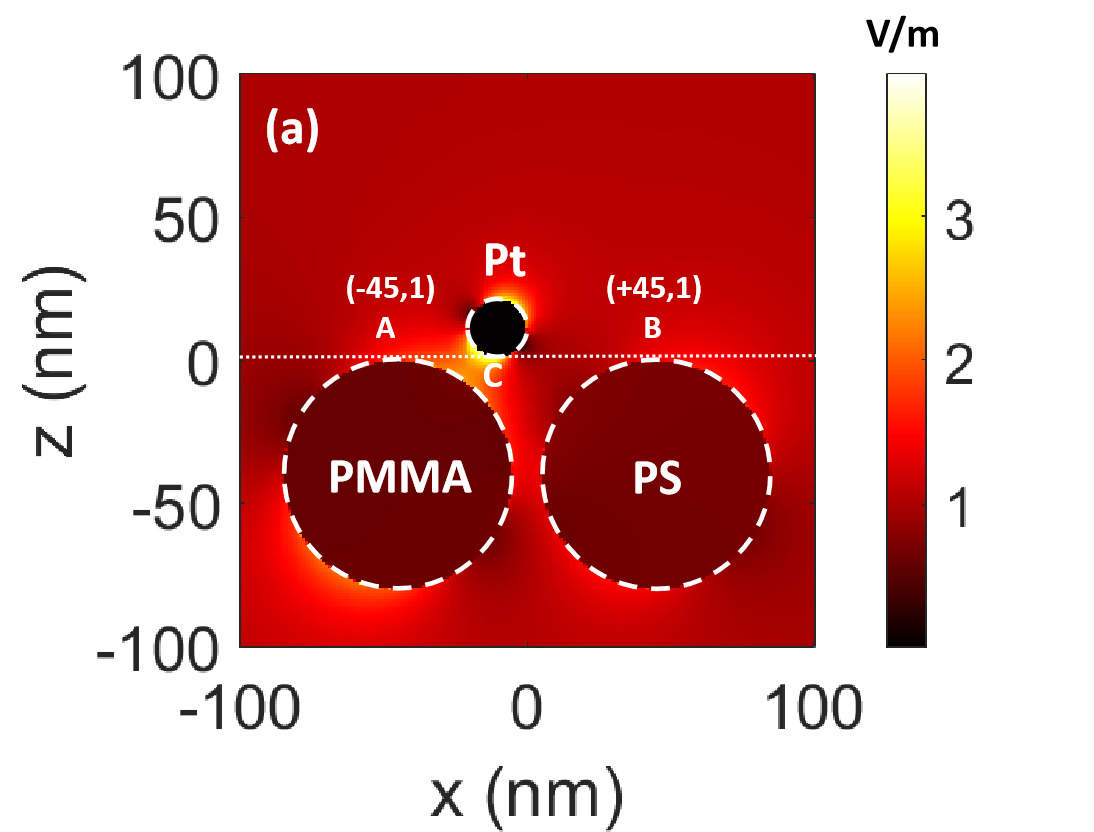}
    \includegraphics[width=0.49\columnwidth]{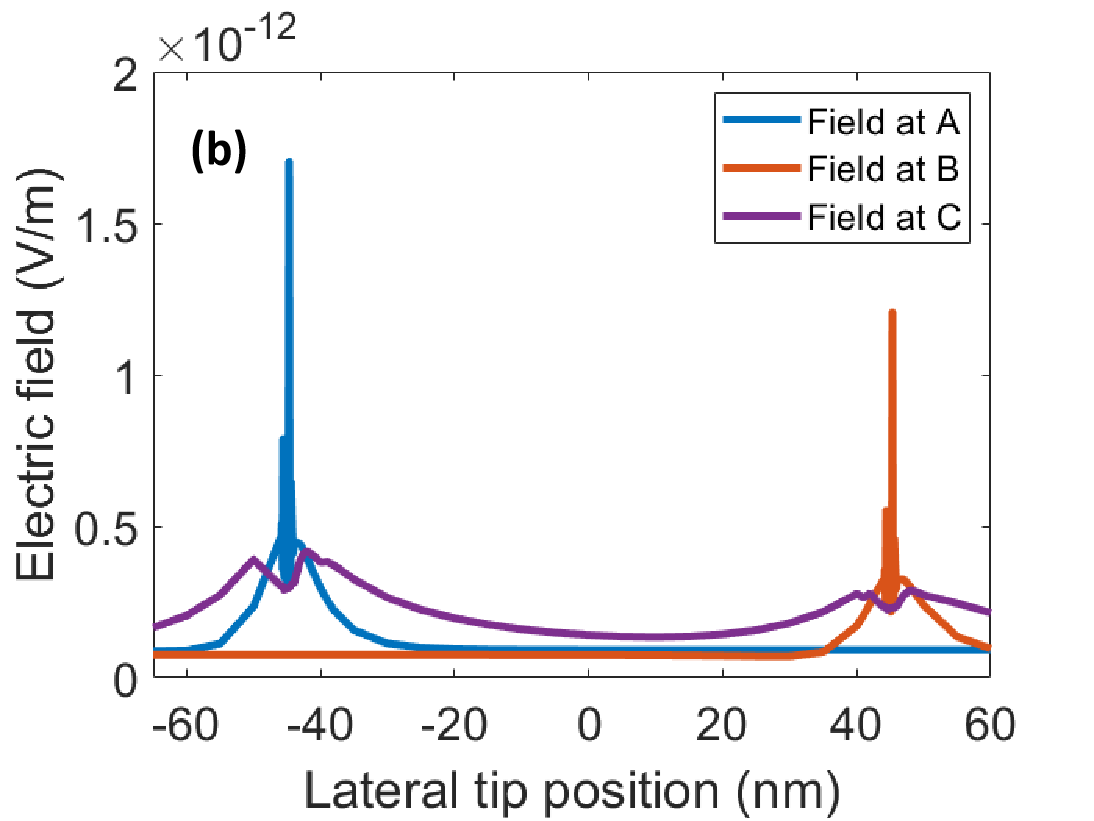}
    \includegraphics[width=0.49\columnwidth]{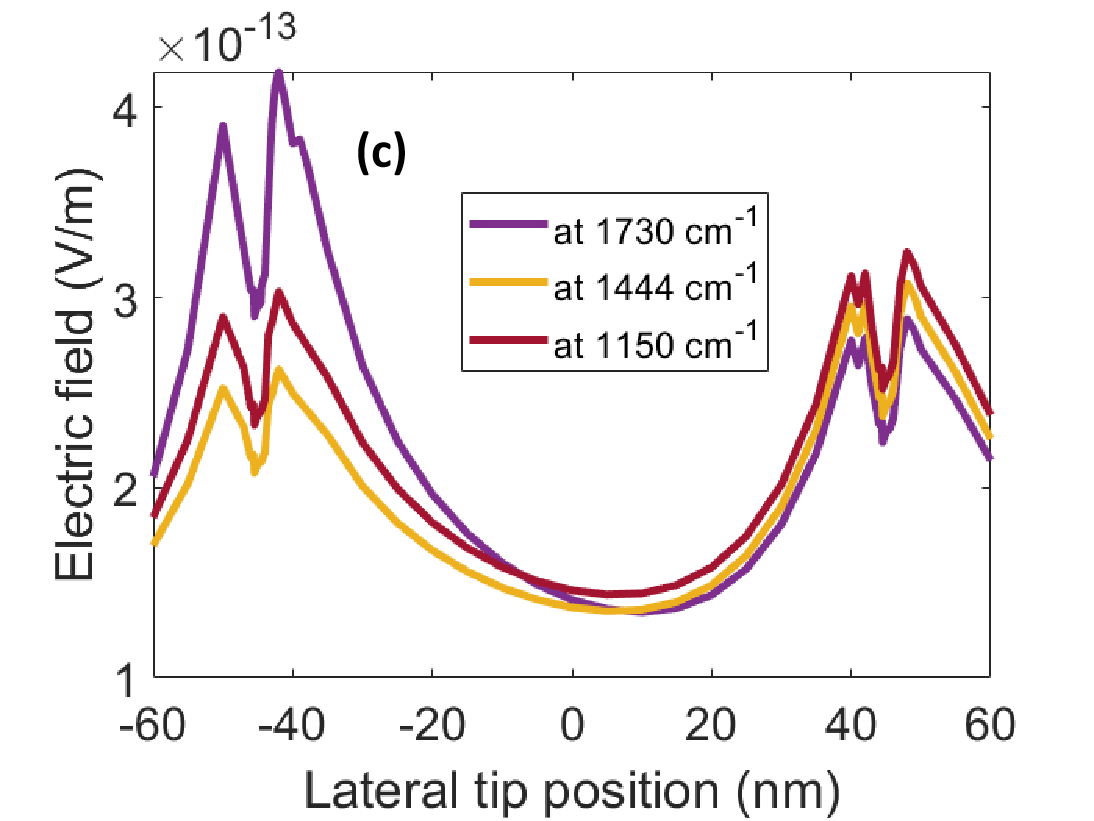}
    \caption{Model experiment: A Pt spherical tip with $r=10$~nm scanning over a PMMA and a PS NP with $R=$~40~nm, each, at 10~nm interdistance in $x$. (a) 2D electric field map presenting the simulation setup for an arbitrary scan position. (b) Electric field strength as a function of the tip position  measured at the three marked positions A, B and C in the system for $\nu=1730$~cm$^{-1}$. (c) Electric field strength obtained in C as a function of the tip movement at three different absorption frequencies $\nu$ of PMMA.}
    \label{fig8}
\end{figure}

Finally, we consider the situation of scanning over two particles of the same size, but different materials, to demonstrate that the simulated probe is able to distinguish between them. In this model setup, two spherical polymer NPs with $R=40$~nm (PMMA and PS) are placed 10~nm apart from each other. A Pt sphere with $r=10$~nm mimicking the AFM tip with oscillation amplitude $A_2=1$~nm is performing a line scan at a fixed height from left to right at contact, i.e., its equilibrium position is set as $z_0=r_0+1$~nm with respect to the top of the NPs, see the marked line in Fig.~\ref{fig8}(a). During the lateral motion of this "tip", the electric field is computed in two fixed positions at $z_0 = r_0 +1$~nm above each NP, marked by A and B, respectively, as well as in a moving position C, below the laterally moving Pt sphere at $z= z_0-1$~nm.

First, we consider $\nu=1730$~cm$^{-1}$, an absorption frequency of PMMA. When the tip approaches the PMMA NP, a strong increase of the field is observed in A and C, see Fig.~\ref{fig8}(b), which results from enhancement in the hybrid system.~\cite{Lukyanchuk2010, Khurgin2024} However, directly above the NPs, a slight reduction in field enhancement is seen. When the tip is moving further to the right, a second increase in the electric field is observed, which is larger than the first one. A similar but weaker field modulation is observed in B and C, when the tip is close to the PS NP. The asymmetric shape of the field enhancement close to the NPs is caused by the offset between the angle of incidence and the oscillation of the tip parallel to the sample normal. Hence, even under ideal circumstances, the overall tip-sample interaction may remain asymmetric due to the incidence angle of the light with respect to the tip oscillation and the optical coupling in the hybrid system leading to non-dipolar field distributions. 

The smaller field enhancement above the PS NP compared to that above the PMMA NP is a result of the choice of illumination frequencies, which match the resonant condition for PMMA only. A similar variation of $E$ is also observed for excitation at two other absorption frequencies of PMMA, see Fig.~\ref{fig8}(c) for the field measured in C. On top of the PMMA NP, the strongest field enhancement is observed for $\nu=1730$~cm$^{-1}$, and the weakest for $\nu=1444$~cm$^{-1}$, which is a result of the different absorption coefficients of PMMA in the three bands. However, the slight intensity variation between the three frequencies on top of the PS NP is caused by differences in the formation of the tip plasmon at the varied $\nu$.~\cite{Huth_antenna2013} Next, we discuss a physical experiment in a similar setting and compare the observations.

\subsection{Experimental PiF contrasts of a spherical PMMA nanoparticle}
\begin{figure}[t!]
    \centering
    \includegraphics[width=\columnwidth]{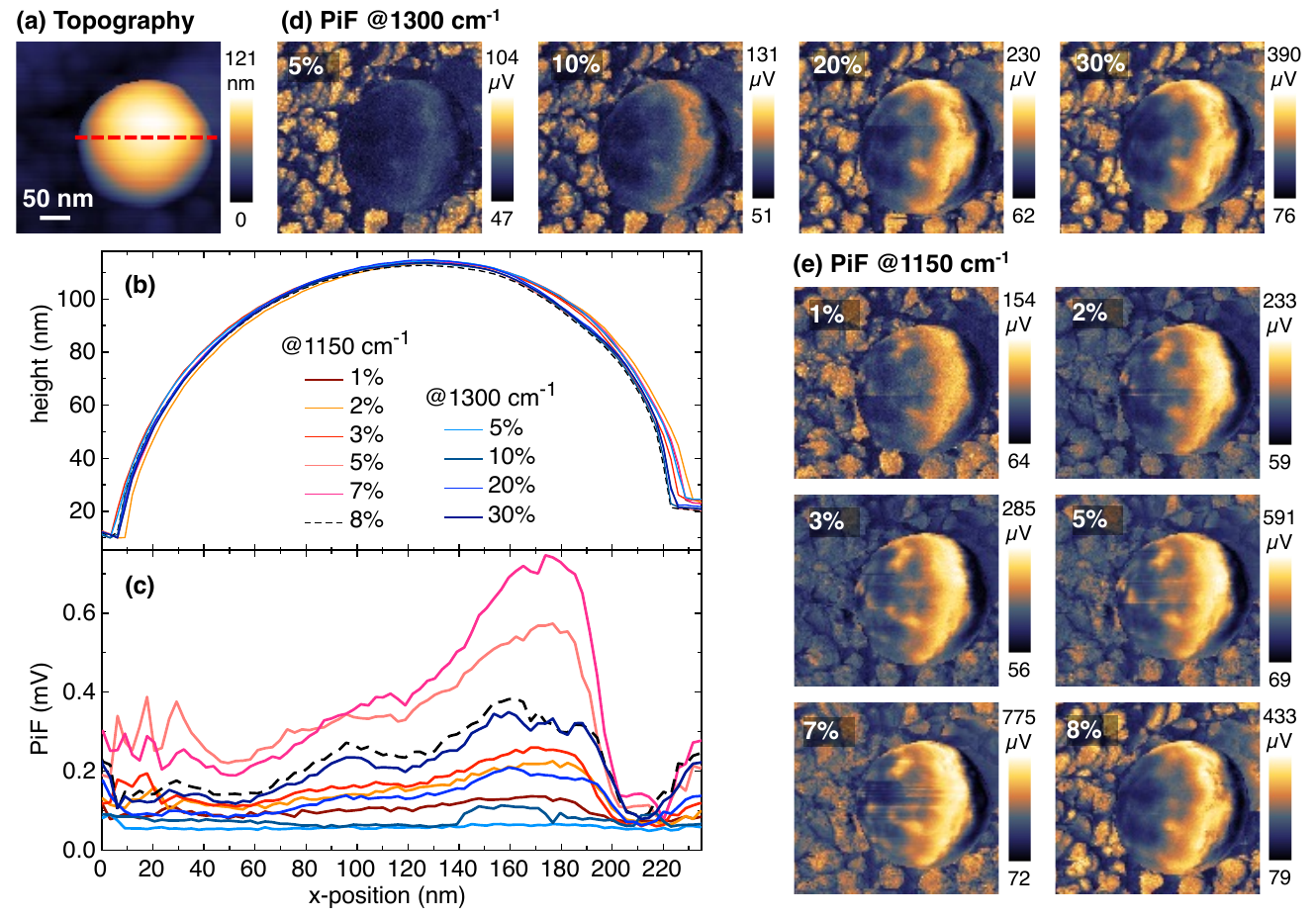}
    \caption{Experimental PiF contrasts of a spherical PMMA NP with $R=50$~nm at varied illumination power: a) AFM topography (@1150 cm$^{-1}$, $I_0=1$\% $I_{\rm{max}}$), b) height profiles and c) PiF intensity profiles across a line in all cropped scan images (red line in a), d) PiF outside of PMMA absorption at 1300~cm$^{-1}$ and e) PiF within PMMA absorption band at 1150~cm$^{-1}$.}
    \label{fig9}
\end{figure}
The experimental evaluation of the PiF contrasts obtained from scanning a spherical PMMA NP with $R=50$~nm using a Pt-coated AFM tip reveals an even stronger asymmetry than obtained from our model, see Fig.~\ref{fig9}. The NP has an apparent width  $w_A \approx$ 200~nm, while its height above the background is $\approx$ 120~nm, which is closer to the size of $100\pm5$~nm given by the manufacturer, see the example of an AFM topography scan ($\nu=1150$ cm$^{-1}$, intensity of illumination $I_0=1$\% of the max intensity $I_{\rm{max}}$ of the QCL power spectrum measured at the tip) and the line profiles of all cropped scans in Fig.~\ref{fig9}b) and c) along the marked line in Fig. \ref{fig9}a). 
A convolution of a NP with the PPP tip shape adds the width $w_{h=50}$ of the tip in $\approx 50$~nm height from the apex on both sides of the NP, while the height of the NP should be reproduced with less error.~\cite{ricci_recognizing_2004} Using the parameters given by the manufacturer, we obtain: 
$w_A = 2R + 2w_{h=50}\approx100$~nm~$+ 2(100\tan{10\degree}+14)$~nm = 163~nm. Taking also the inclination of the tip into account, the observed $w_A \approx$ 200~nm is realistic. 

\begin{figure}[t!]
    \centering
    \includegraphics[width=\columnwidth]{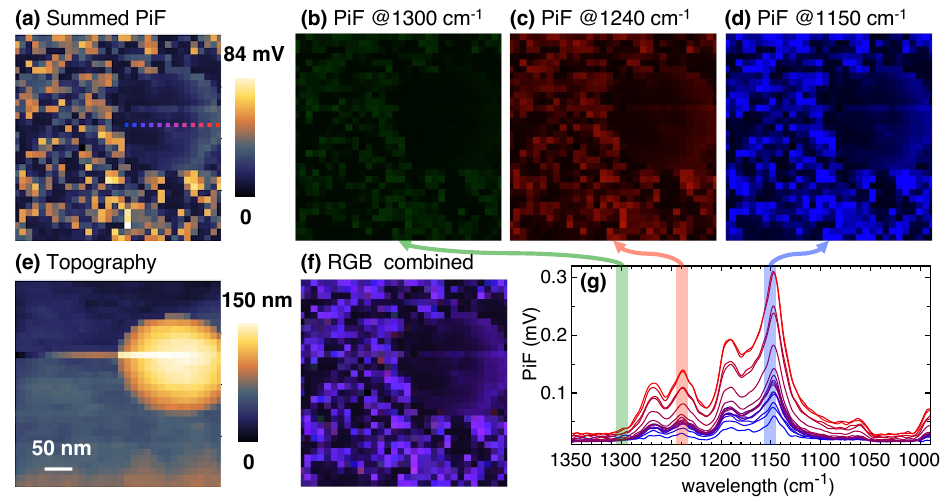}
    \caption{Experimental PiF contrasts of the NP in the spectral range of 990 - 1350~cm$^{-1}$. a) PiF summed over the spectral range, b-d) PiF contrasts at selected spectral bands: 1300, 1240 and 1150 ~cm$^{-1}$, respectively, e) AFM topography, f) combined RGB image of (b-d) and e) PiF-IR spectra in matching color scale along the line in (a) with the spectral bands in (b-d) marked.}
    \label{fig10}
\end{figure}

The NP is surrounded by material flakes from fabrication, which absorb also at 1300~cm$^{-1}$ outside of the PMMA aborption bands and thus chemically differ from the material of the NP, see Fig.~\ref{fig9}d). Unfortunately these flakes easily attach to the scanning AFM tip. As a result, horizontal stripes are seen in several of the scans in Fig.~\ref{fig9}d) and e) and also in the hyperspectral scan in Fig.~\ref{fig10}. In particular, such a flake attached to the tip seems to have caused the decreased PiF intensity found for the highest $I_0=8$\% for $\nu=$ 1150~cm$^{-1}$ (dashed height profile in Fig.~\ref{fig9}c) and corresponding PiF contrast in Fig.~\ref{fig9}e)). To enable the evaluation of the impact of the varied $I_0$ and $\nu$ on the NP shape, the height profiles have been vertically shifted to a common minimum position at the left side of the NP, which involved a shift of $-26$~nm and of ${-1}$ to ${-5}$~nm for the scan at $I_0=8$\% and for all other scans, respectively. This agrees with the assumption of an attached flake during the scan of the former. In PiF-IR, attached molecules and soft coatings strongly impact the tip-induced near-field enhancement,~\cite{li_tip-enhanced_2020} which in case of the here used $\nu=$ 1150~cm$^{-1}$ resulted in a considerable reduction of the detected PiF.

By comparing the results obtained at varied $I_0$ (in the range of a few 100 µW) in the resonant condition at $\nu=$ 1150~cm$^{-1}$ with those for the non-resonant condition at $\nu=$ 1300~cm$^{-1}$, we can discriminate the effect of the photo-induced thermal expansion of the PtIr-coated tip from the combined effect of both, the thermal expansion of the tip and the near-field enhanced absorption causing the thermal expansion in the NP.
In the non-resonant condition at weak intensity ($I_0=5$\%), the NP shows negligible absorption (light blue line in Fig.~\ref{fig9}e)). This changes with increasing $I_0$. At $I_0=20$\%, the PiF profile is comparable to that for $I_0=2$\% in the resonant condition at $\nu=$ 1150~cm$^{-1}$, compare the royal blue with the red line in Fig.~\ref{fig9}e), respectively. However, the shape of the profiles slightly differ: in the non-resonant condition, the PIF peaks at $x\approx 160$~nm, which is closer to the top of the sphere ($x\approx 120$~nm) than the PiF profile maximum ($x\approx 180$~nm) in the resonant condition. At high $I_0$, there is a considerable contribution of the heat and the thermal expansion of the metallic tip itself due to its broad plasmonic resonance at mid-IR illumination frequencies.\cite{Huth_antenna2013} In the resonant condition, the electric field coupling between the absorbing nanostructure and the tip plasmon dominates the geometry of the PiF profile across the nanostructure at least for moderate $I_0$. The, in general, slightly wider height profiles for $\nu=1150$~cm$^{-1}$ compared to those for $\nu=1300$~cm$^{-1}$ in Fig.~\ref{fig9}b) match to this consideration. For the resonant condition, there seems to be an additional expansion of the NP in direction of the inclined tip. However, this effect could be also caused by small mismatches in cropping the scan positions due to the discrete scan positions, which could be improved by increasing the scan resolution in further studies. Nevertheless, the observation of an intensity dip on top of the NP and a stronger enhancement at the side opposite to the incident light (at $\approx 80\degree$ from the left), in general, matches the above presented prediction using a rather simple simulated setup. Furthermore, in our simplified model we assumed a constant height of the tip with respect to the substrate during the later scan. In the experiment, the tip follows the height structure in the sample and the PiF profile is affected by the convolution of the nanostructure with the inclined tip shape and resulting effects on the tip-induced near-field enhancement. 

The spatial contrast of the PiF at varied $\nu$ is further demonstrated in the analysis of a hyperspectral scan in the range between 990 and 1350~cm$^{-1}$, see Fig.~\ref{fig10}. From the hyperspectrum, PiF contrasts in three spectral bands $\nu=$ 1300, 1240, and 1300~cm$^{-1}$ of width $=10$~cm$^{-1}$ have been selected, homogenized to the same intensity range (0-5 mV) and presented as RGB contrasts in green, red and blue, respectively. The combined RGB image (Fig.~\ref{fig10}f) confirms the absorption in the red and blue channels. In agreement with the above discussed PiF profiles, the PiF-IR spectra across the marked line on the NP (Fig.~\ref{fig10}a.) vary in their overall intensity. The spectra are plotted in corresponding colors to the dots in the line in Fig.~\ref{fig10}g). The highest intensity is observed at the right end of the line, which is close to the right end of the NP (red). The intensity variation along the marked line is not completely linear over this spectral range. In particular, the spectrum at the right end of the line shows a slight intensity increase at $\approx1270$~cm$^{-1}$, which might be due to some contribution from the absorption of the flakes surrounding the NP. Nevertheless, there is no strong deviation from a linear dependence on intensity over the spectral range. The intensity variation, in general, matches the PiF profiles in Fig.~\ref{fig9}e). This confirms a dominating impact of the direction of the illuminating light beam on the detected PiF in case of this isotropically absorbing PMMA NP. For ansiotropically absorbing nanostructures, an additional impact of the material alignment in respect to the local polarization properties of the plasmonic near-field of the tip is expected.~\cite{luo_intrinsic_2022} 

\section{Conclusion}
We presented a comprehensive insight into the interplay of light absorption, coupling of the optical fields between the metallic AFM tip and the nanostructure and the resulting photothermal expansion which leads to a change in the attractive van der Waals force acting between the tip and the nanostructure and results in the detected PiF. We analyzed the tip-sample interaction force $F_{\text{ts}}$ by calculating the local temperature change due to the optical absorption and the corresponding thermal expansion for different tip apexes and materials. We investigated varying illumination conditions such as the polarization of the incoming light and its effect on the resulting forces depending on different tip movements. We modeled a simplified 2D scan of a nanostructure incorporating a variation in the chemical composition and demonstrated that within the framework of our simulation it is possible to discriminate the chemical composition of the nanostructure due to the change in the acting forces. Although we used a simplified model which neglects the lateral thermal expansion of the nanostructure and the expansion of the tip, essential features of the experimentally obtained PiF contrast from scanning a spherical PMMA NP at resonant illumination frequencies could be reproduced. Based on the developed model, further investigation can be carried out on more realistic surface morphologies such as asymmetric NPs and rough surfaces including nonlocal proximity effects~\cite{David2011a, Christensen2014} or nonlinear effects due to intense local fields. 
Furthermore, the developed model can be extended to include also an improved heat diffusion model in the tip-sample junction together with the incorporation of lateral thermal expansion in the sample and the tip.~\cite{Ritz2023_tipvibrations} This will refine the prediction of the response of nanostructure materials in PiF-IR, which is a further required step for a quantitative analysis of the chemical composition in nanostructure materials and it will help to exploit the promising potential of the unprecedented spatial and high spectral resolution of PiF-IR for high-resolution chemical characterization in the Materials and Life Sciences.

\section*{Acknowledgments}
This research received support from the \textit{Profillinie LIGHT}-program of the Friedrich-Schiller University Jena through funding of the project “Photo-induced thermal expansion of rough surface morphologies (pintXsum)”. The authors thank the Leibniz-Institute of Photonic Technology for providing access to the VistaScope instrument, for which financial support of the European Union via the Europ\"{a}ischer Fonds für Regionale Entwicklung (EFRE) and the “Th\"{u}ringer Ministerium für Bildung Wissenschaft und Kultur (TMBWK)” (Project: 2018 FGI 0023) is highly acknowledged. We thank Rainer Heintzmann, Leibniz Institute of Photonic Technology in Jena and Friedrich Schiller University Jena, for helpful scientific discussions.
%

\section*{Data availability}
Raw data sets are available on Zenodo:~\cite{tauber_photothermal_2024}  https://doi.org/10.5281/zenodo.14208617 


\providecommand{\latin}[1]{#1}
\makeatletter
\providecommand{\doi}
  {\begingroup\let\do\@makeother\dospecials
  \catcode`\{=1 \catcode`\}=2 \doi@aux}
\providecommand{\doi@aux}[1]{\endgroup\texttt{#1}}
\makeatother
\providecommand*\mcitethebibliography{\thebibliography}
\csname @ifundefined\endcsname{endmcitethebibliography}  {\let\endmcitethebibliography\endthebibliography}{}



\section*{Supplementary material}

# Supporting Information:

# Photothermal Expansion of Nanostructures in Photo-induced Force Microscopy

Shohely Tasnim Anindo,[†,‡] Daniela Täuber,[¶,§] and Christin David[*,†,∥]

†*Institute of Condensed Matter Theory and Optics, Friedrich-Schiller-Universität Jena, Max-Wien-Platz 1, 07743 Jena, Germany*

‡*Abbe Center of Photonics, Albert-Einstein-Straße 6, 07745 Jena, Germany*

¶*Institute of Physical Chemistry, Friedrich-Schiller-Universität Jena, Helmholtzweg 4, 07743 Jena, Germany*

§*Leibniz Institute of Photonic Technology, Albert-Einstein-Straße 9, 07745 Jena, Germany*

∥*University of Applied Sciences Landshut, Am Lurzenhof 1, 84036 Landshut, Germany*

E-mail: christin.david@haw-landshut.de

### Abstract

The Supporting Information provides technical details on the computational method as well as material parameters used in the calculations.

# Boundary Element Method

There are several methods to solve Maxwell's equation for computing the electric field inside and outside a medium such as Finite Element Method (FEM), Discrete Dipole Approximation (DDA), Boundary Element Method (BEM) and Finite Difference Time Domain

(FDTD).[S1,S2] BEM expresses the electromagnetic field scattered by a NP in terms of boundary charges and currents. By applying the standard boundary conditions for the continuity of the parallel components of the electric and magnetic fields, a system of surface-integral equations is formed. This method assumes a homogeneous environment where structures with isotropic dielectric functions are separated by defined surfaces discretized with a set of points that lead to a large system of surface integrals. This discretization transforms the problem into a set of linear equations which can be solved numerically using standard linear algebra techniques.[S2,S3] The BEM technique is well suited for most plasmonics applications involving metallic NPs placed into a dielectric background. Its advantage is that faster simulations with a more reasonable memory demand are achieved by discretizing only the boundaries between particle and environment, rather than the entire volume as with, e.g. FDTD.

## Importance of retardation

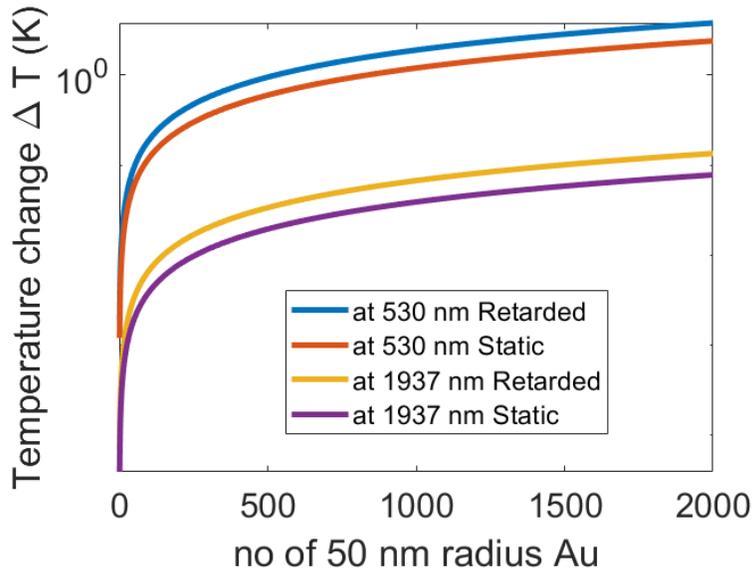

Figure S1: Showing the temperature change as a function of the number of Au NPs with a radius of 50 nm comparing the results for a quasi-static and retarded calculation using the electric field amplitude $54 \times 10^3$ V/m.

In the simplistic quasi-static approach, when the particle size is assumed to be much smaller than the excitation wavelength an instantaneous interaction where the speed of light is infinite is used. This assumption results in decoupling the electric field and magnetic fields in Maxwell's equations. The magnetic response is negligible at optical frequencies leaving only electric field to be considered. The electric field is then expressed as the gradient of the scalar electric potential $\mathbf{E} = -\Delta\phi$. Conversely, taking the finite speed of light into account leads to retardation effects. As a result, both electrostatic and vector potential are needed to compute the electric field, $\mathbf{E} = ik\mathbf{A} - \Delta\phi$, where $\mathbf{H} = \nabla \times \mathbf{A}$.[S4,S5]

Fig. S1 represents the total heat generated by a number of individual Au NPs of 50 nm radius illuminated at a wavelength $\lambda = 530$ nm (18867.9 cm$^{-1}$) and $\lambda = 1937$ nm (5162.6 cm$^{-1}$) with an electric field amplitude of $54 \times 10^3$ V/m. At this particle size and for distances between regularly distributed NPs such that their coupling becomes sizable, there is a significant difference in results neglecting retardation vs. full field calculations, see Fig. **??**. The temperature values provided by the retarded method are always larger than from the static measurement. For 2000 Au NPs, at $\lambda = 530$ nm, the temperature change is 3.8 K

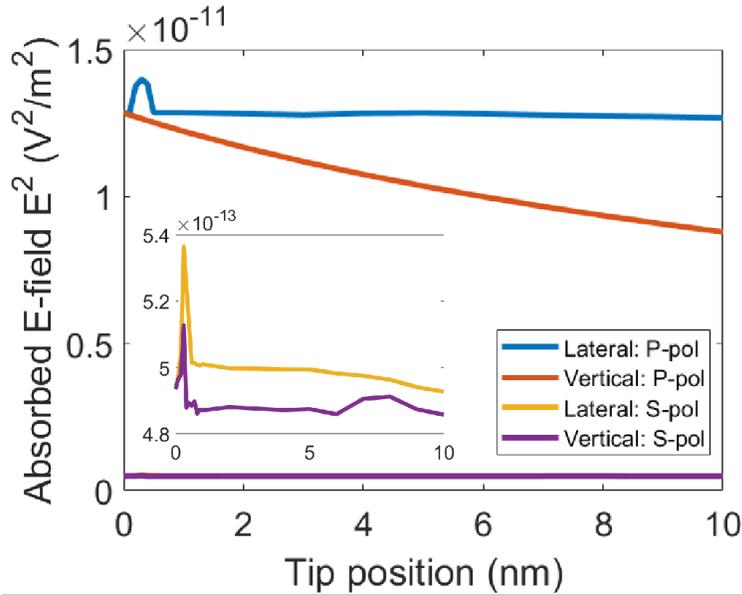

Figure S2: Absorbed field intensity $I \sim E^2$ ($V^2/m^2$) for $\nu = 1492$ cm$^{-1}$ by a PS nanosphere with $R = 50$ nm as a function of the lateral and vertical displacement of a tip with $r = 10$ nm for both states of polarization.

with retardation and amounts to 2.4 K without.

Considering the lateral and vertical displacement of polymer NPs illuminated within their respective absorption bands in Fig. S2 we observe, that for P-polarization, the electric field absorbed by the NP increases up to an optimal distance $z \approx 0.3$ nm in both cases and decreases afterwards. This decrease is faster with the vertical shift were the plasmon coupling is lost more quickly than in the lateral movement. In contrast, the change in the absorbed electric field due to the tip position is reduced by two orders of magnitude for S-polarization due to minimal coupling effects.

## Surface discretization

The boundary discretization is a vital instrument in BEM calculations. For spherical NPs, the upper limit for surface points is 1444 in the provided triangular meshing of the MNPBEM toolbox.[S4,S6] Metals and other dielectric materials with a permittivity that exhibits large imaginary parts rapidly produce inaccurate electric field values for a lower number of boundary elements as the contrast in the refractive index to the environment is larger than in glassy materials. The computational time increases significantly with the number of boundary elements. For a resolution of 1 nm, a single script computing the electric field in the environment for two spherical NPs each with 1444 surface points for a single wavelength excitation takes several days. With BEM, the NP's response to the field can be analyzed in 2D cross-sections, which takes comparatively less time per single spectrum (a few hours for 5000 parameterization points.[S2]) On the other hand, the field at individual points within or around the NP allows studying how the electromagnetic field varies spatially. This localized field analysis reveals the detailed distribution and intensity of the electric and magnetic fields at specific locations, which is crucial for understanding phenomena like hot spots, plasmonic effects, and near-field interactions.

Table S1: Fundamental properties of materials.

| | Operating wavelength $\lambda$ | Operating frequency $\nu$ | Dielectric constant $\epsilon$ | Thermal expansion coefficient $\alpha$ | Effective thermal conductivity $\kappa_{eff}$ | Density $\rho$ | Heat capacity $C$ | Thermal diffusivity $D$ |
|---|---|---|---|---|---|---|---|---|
| | nm | cm$^{-1}$ | | 10$^{-6}$/K | W/Km | kg/m$^3$ | J/K] | m$^2$/s |
| Au | 530 | | -4.55+2.46i | 14 | 314 | 19450 | 125 | 1.27e-4 |
| | 1937 | | -189.04+25.36i | | | | | |
| PMMA | 5780.35 | 1730 | 2.18+2.85i | 76 | 0.19 | 1180 | 1466 | 3.0e-7 |
| | 6925.21 | 1444 | 2.13+0.25i | | | | | |
| | 7692.31 | 1300 | 1.92 + 0.08i | | | | | |
| | 8695.65 | 1150 | 2.36+1.13i | | | | | |
| PS | 6702.42 | 1492 | 2.46+ 0.23i | 70 | 0.17 | 920 | 1300 | 2.5 e-7 |
| Pt | 1770.70 | | -13.98+77.03i | 8.9 | 71 | 21400 | 133 | 2.52e-5 |



\end{document}